\documentclass[amssymb,aps,amsmath,floatfix,prl,twocolumn,superscriptaddress]{revtex4-2}
\usepackage[utf8]{inputenc}

\usepackage{mathptmx}
\usepackage[T1]{fontenc}
\usepackage{graphicx}
\usepackage{amsmath,amssymb,amsfonts, MnSymbol}
\usepackage{mathrsfs}
\usepackage{bm}
\usepackage{hyperref}
\usepackage{bbold}
\usepackage{xcolor}
\usepackage{tikz}

\newcommand{\algn}[1]{\begin{align} #1 \end{align}}
\newcommand{\sbeqs}[1]{\begin{subequations} #1 \end{subequations}}

\newcommand{\ve}[1]{\boldsymbol{#1}}
\newcommand{\nn}{\nonumber}
\newcommand{\ee}{\ensuremath{\text{e}}}
\newcommand{\ed}{\ensuremath{\text{d}}}

\newcommand{\dd}[1]{\ensuremath{\tfrac{\text{d}}{\text{d} #1}}}

\newcommand{\mc}[1]{\ensuremath{\ms{#1}}}
\newcommand{\ms}[1]{\ensuremath{\mathscr{#1}}}
\newcommand{\mbb}[1]{\ensuremath{\mathbb{#1}}}
\newcommand{\kb}{\ensuremath{k_\text{B}}}
\newcommand{\eqnlab}[1]{\label{eq:#1}}

\newcommand{\figlab}[1]{\label{fig:#1}}
\newcommand{\eqnref}[1]{\eqref{eq:#1}}
\newcommand{\Eqnref}[1]{Eq.~\eqref{eq:#1}}
\newcommand{\Eqsref}[1]{Eqs.~\eqref{eq:#1}}

\newcommand{\figref}[1]{\ref{fig:#1}}
\newcommand{\Figref}[1]{Fig.~\ref{fig:#1}}
\newcommand{\Figsref}[1]{Figs.~\ref{fig:#1}}

\begin{document}
\title{Exponential Change of Relaxation Rate by Quenched Disorder}

\author{Jan Meibohm}
\affiliation{Technische Universit\"at Berlin, Institut f\"ur Theoretische Physik, Hardenbergstraße 36, 10623 Berlin, Germany}
\author{Sabine H. L. Klapp}
\affiliation{Technische Universit\"at Berlin, Institut f\"ur Theoretische Physik, Hardenbergstraße 36, 10623 Berlin, Germany}

\begin{abstract}
	We determine the asymptotic relaxation rate of a Brownian particle in a harmonic potential perturbed by quenched Gaussian disorder, a simplified model for rugged energy landscapes in complex systems. Depending on the properties of the disorder, we show that the mean and variance of the asymptotic relaxation rate are non-monotonous functions of the parameters for a broad class of disorders. In particular, the rate of relaxation may either increase or decrease exponentially compared to the unperturbed case. This implies that disorder may, depending on its properties, both significantly speed up \textit{and} slow down relaxation. In the limit of weak disorder, we derive the probability distribution of the asymptotic relaxation rate and show that it is Gaussian, with analytic expressions for the mean and variance that feature universal limits.  Our findings indicate that controlled disorder may serve to tune the relaxation speed in complex systems.
\end{abstract}
\maketitle
\section{Introduction}
Thermal relaxation is a fundamental process in statistical physics that is of relevance across a wide range of disciplines in science and engineering. However, most of our current understanding of thermal relaxation is limited to local-equilibrium situations, where linear-response techniques apply. Genuine far-from-equilibrium relaxation, by contrast, lacks a systematic treatment, yet it offers a variety of fascinating, anomalous phenomena.
Well-known examples include the Mpemba effect~\cite{Mpe69,Bai19}, memory phenomena in glasses~\cite{Arc22} such as ageing or the Kovacs effect~\cite{Kov63,Cal06}, coarsening in phase-ordering kinetics~\cite{Bra02}, or finite-time dynamical phase transitions~\cite{Vad24}.

Anomalous-relaxation effects that are observable at macroscopic scales typically occur in interacting many-body systems whose dynamics are difficult to describe. Recent progress in the understanding of such effects has therefore focused on simplified, mesoscopic models that often involve only a few degrees of freedom in contact with a heat bath. Such simplifications enable comprehensive theoretical and experimental descriptions of, e.g., mesoscopic analogues of the Mpemba effect~\cite{Lu17,Kli19,Kum20,Che21} and of similar relaxation asymmetries~\cite{Lap20,Mei21b,Vu21,Iba24}, as well as of dynamical phase transitions~\cite{Mei22a,Mei23b,Tez23b} in the relaxation process. Relaxation anomalies are also studied in simplified quantum systems, both open~\cite{Cha23,Car21,Wu24,Jos24,Ryl24} and isolated~\cite{Pta24}, and they have motivated the search for ways to optimise relaxation~\cite{Def14,Tez23a,Gue23}, for instance by using pre-heating strategies~\cite{Gal20,Pem21,Pem24}, by adding a nonequilibrium drive~\cite{Hwa93,Suw10,Ich13,Cog21,Die23}, or by designing optimal manipulation protocols~\cite{Mar16,Gue23,Ray23}.

In addition to their size, a major difference between mesoscopic model systems and the real-life many-body systems they are motivated by, is that real-life systems are often subject to static, ``quenched'' disorder, caused by, e.g., the presence of impurities. Quenched disorder has been observed to slow down relaxation in the random-field Ising model~\cite{Nat88}, spin glasses~\cite{Cug94,Bai19}, fluids in mesoporous materials~\cite{Kla01,Woo03,Det03}, and other systems~\cite{Bir15,Dua21}. However, a detailed understanding of the effects of disorder on relaxation in complex systems remains elusive.

In this Letter, we introduce a simple model that serves as a test bed for the effect of quenched disorder on the rate of thermal relaxation. The model is based upon a single Brownian particle in a one-dimensional harmonic potential, a paradigm for the study of classical relaxation phenomena. Quenched disorder is introduced by adding a static Gaussian random potential with a predefined spatial correlation function. For each disorder realisation, the full potential is more complex than the unperturbed one, comprising, for example, additional local minima and maxima.

The quenched disorder may represent experimental imperfections in optical traps, e.g., in the presence of speckles~\cite{Dri10,Vol14,Han12,Bew16a,Zun22}. More generally, the model mimics the effect of rugged energy landscapes on relaxation in complex many-body systems. Related models have been studied in the context of spin glasses~\cite{Fra94} and of disordered quantum systems, e.g., in theoretical~\cite{Bho10,Hsu18,Hsu20,Schu22} and experimental~\cite{Dri10} studies of Bose-Einstein condensates.

We analyse the model and show that the disorder properties have profound impacts on the asymptotic relaxation rate of the Brownian particle. In particular, disorder can both exponentially decrease \textit{and} increase the average rate of relaxation. This exponential change of relaxation rate by quenched disorder is in contrast to other relaxation anomalies such as the Markovian Mpemba effect~\cite{Lu17} or relaxation asymmetries~\cite{Lap20}, whose effects are typically sub-exponential (see Refs.~\cite{Kli19,Die23}, however, for notable exceptions). The exponents associated with the disorder-induced change are shown to be Gaussian distributed at weak disorder, with mean and variance computed explicitly.

Our findings imply that correlated disorder, whose effects are usually tried to be mitigated, may serve to tune the speed of relaxation by manipulating the disorder properties.
\section{Model}
We consider thermal relaxation of a particle immersed in a fluid at inverse temperature $\beta = (\kb T)^{-1}$ and subject to a disordered energy landscape. The latter consists of a harmonic potential of stiffness $\alpha$, amended with a Gaussian disorder field of magnitude $\zeta$. The motion of the particle is overdamped with the damping time scale~$\tau = (\mu\alpha)^{-1}$, where $\mu$ denotes the mobility. For a single disorder realisation, the expectation value $\langle f(x)\rangle$ for an arbitrary function $f$ of the dimensionless particle position $x$ obeys
\algn{\eqnlab{fpeqn}
	\dd{t} \langle f(x)\rangle =\langle\ms{L}^\dagger f(x) \rangle\,, \quad \ms{L}^\dagger = \left[-W'(x) + \partial_x\right]\partial_x\,,
}
where $\langle\ldots\rangle$ is the average over the thermal noise and $\ms{L}^\dagger$ denotes the adjoint of the (non-Hermitian) Fokker-Planck operator $\ms{L}$~\cite{Ris89}. In this dimensionless formulation, the spatial coordinates are measured in units of the thermal length scale $\ell_T = 1/\sqrt{\alpha\beta}$, so that the magnitude of equilibrium fluctuations of $x$ at vanishing disorder equals unity. Time $t$ is measured in units of $\tau$. The potential $W(x)$ in \Eqnref{fpeqn} is decomposed as
\algn{\eqnlab{potential}
	W(x) = \frac{x^2}2 + \zeta_T V\left(\frac{x}{\kappa}\right)\,.
}
The first term in \Eqnref{potential} corresponds to the harmonic part of the potential. 
The second term represents the disorder, consisting of a time-independent Gaussian random field $V$ multiplied by the dimensionless coupling $\zeta_T \equiv \beta\zeta$. The mean and spatial correlations of $V$ are given by
\algn{\eqnlab{meanvar}
	\langle V(x) \rangle_V = 0\,,\quad \langle V(x) V(y) \rangle_V = C\left(x-y\right)\,,
}
where $\langle\ldots\rangle_V$ denotes the disorder average. The argument of $V$ in \eqnref{potential} is rescaled by $\kappa\equiv \ell_c/\ell_T$, denoting its correlation length $\ell_c$ in units of $\ell_T$. The correlation function $C(x)$ has the properties
\algn{\eqnlab{corrprop}
	C(0) = 1\,,\quad C(-x) = C(x)\,,\quad \int_{0}^\infty\!\!\!\ed x\,\, C(x) = 1\,.
}
The first two properties refer to the normalisation and symmetry of $C$, while the third one ensures a finite correlation length.

The shape of $C$ and the magnitude of $\kappa$ dictate the properties of realisations of the full potential $W$. In particular, the number of times $C(x)$ is differentiable at $x=0$ determines the smoothness of realisations of $V$. Figure~\figref{pot_reals}(a) shows three correlation functions, Gaussian~\footnote{The intensity of laser speckles is typically Gaussian correlated~\cite{Goo07}. Therefore, Gaussian correlation functions are relevant for speckle-perturbed laser traps.}, sinc, and exponential, that satisfy the conditions in \Eqnref{corrprop}. While two of these (Gaussian and sinc) are smooth at $x=0$, the exponential correlation function $C(x)=\exp(-|x|)$ is not. 

Figure~\figref{pot_reals}(b) shows the corresponding realisations of $W$ for varying $\kappa$. We observe that when $C$ is smooth, then so are the realisations of $W$. In this case, larger $\kappa>1$ deforms the harmonic potential, but leaves its single-well shape intact. Smaller $\kappa<1$, by contrast, leads to the formation of multiple local minima of $W$, as mentioned in the introduction. The formation of local, metastable minima is expected to generally slow down relaxation due to particle trapping.

For $C(x)=\exp(-|x|)$, whose second derivative diverges at $x=0$, $W$ is erratic and exhibits an infinite number of local minima, independently of the magnitude of $\kappa$. These properties will prove instrumental for understanding the relaxation behaviour.
\begin{figure}
	\includegraphics{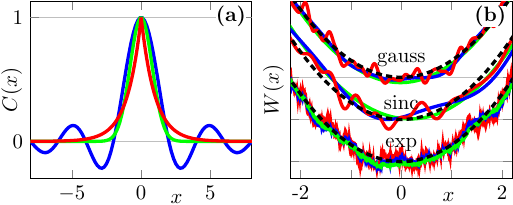}
	\caption{(a) Gaussian $C(x)= \exp(-\pi x^2/4)$ (green), sinc $C(x) = 2\sin(\pi x/2)/\pi x$ (blue), and exponential $C(x)=\exp(-|x|)$ (red) correlation functions. (b) Realisations of $W(x)$ for different $C(x)$ and $\kappa = 0.1$ (red), $\kappa = 0.5$ (blue), and $\kappa = 2$ (green). }\figlab{pot_reals}
\end{figure}

For long times, the relaxation of $\langle f(x)\rangle$ simplifies to~\cite{Ris89}
\algn{\eqnlab{pxtlong}
	\langle f(x)\rangle = \langle f(x)\rangle^\text{eq} + \ee^{-\lambda_1 t} c_1  + \ee^{-\lambda_2 t} c_2  + \ldots\,,
}
where $\lambda_n$ are the eigenvalues of $\ms{L}^\dagger$ with $0=\lambda_0<\lambda_1<\ldots$, $\langle \ldots\rangle^\text{eq}$ denotes an average with respect to the equilibrium distribution $P^\text{eq}(x)= Z^{-1}_\text{eq} \ee^{-W(x)}$ with partition function $Z_\text{eq}$, and $c_n$ are constants that depend on the initial probability distribution and on $f$. The eigenvalues $\lambda_n$ are obtained from the eigenvalue equation
\algn{\eqnlab{lev}
	\ms{L}^\dagger l_n(x) = -\lambda_n l_n(x)\,,
}
featuring the eigenfunctions $l_n$ of $\mc{L}^\dagger$.
\section{Relative relaxation rate}
We observe from \Eqnref{pxtlong} that for long times, the rate of relaxation toward $\langle f(x)\rangle^\text{eq}$ is exponential and dominated by the smallest non-vanishing eigenvalue $\lambda_1$, because $-t^{-1}\ln\left|\langle f(x)\rangle-\langle f(x)\rangle^\text{eq}\right| \sim \lambda_1$ for $t\gg1/(\lambda_2-\lambda_1)$ and for a given $V$. The immediate impact of $V$ on the asymptotic relaxation rate is determined by the relative relaxation rate
\algn{\eqnlab{lam1}
	\Delta\lambda_1\sim-\frac1t\ln\left|\frac{\langle f(x)\rangle-\langle f(x)\rangle^\text{eq}}{\langle f(x)\rangle_{\zeta_T=0}-\langle f(x)\rangle^\text{eq}_{\zeta_T=0}}\right|\,,
}
where $\langle f(x)\rangle_{\zeta_T=0}$ and $\langle f(x)\rangle^\text{eq}_{\zeta_T=0}$ denote the expectation values of $f(x)$ in the unperturbed, harmonic potential at finite time and at equilibrium, respectively.

Since $V$ is a random field, $\Delta\lambda_1$ is a random variable that depends on the disorder realisation. When $\Delta\lambda_1>0$ for a given $V$, relaxation occurs exponentially faster than in the unperturbed system. Conversely, it occurs exponentially slower for $\Delta\lambda_1<0$.
\subsection{Numerical simulations}
We first study the mean and variance of $\Delta\lambda_1$ by numerically solving \Eqnref{lev} for a large number of disorder realisations. Figure~\figref{phasediag}(a) shows the mean $\langle\Delta\lambda_1\rangle_V$ as function of $\kappa$ and $\zeta_T$ for Gaussian-correlated disorder. We observe that $\langle\Delta\lambda_1\rangle_V$ is negative for small $\kappa$, indicating slower relaxation. This agrees with the expectation that additional minima in $W$ lead to particle trapping. Unintuitively, however, upon increasing $\kappa$, $\langle\Delta\lambda_1\rangle_V$ changes sign at the dotted line and becomes \textit{positive} for all measured $\zeta_T$~\footnote{As we explain later, $\langle\Delta\lambda_1\rangle_V$ behaves as $\langle\Delta\lambda_1\rangle_V\propto\zeta_T^2$ for $\zeta_T\ll1$. Therefore, the change of sign of $\langle\Delta\lambda_1\rangle_V$ is only marginally reflected by the colour coding in \Figref{phasediag}(a) for small $\zeta_T$.}. This faster relaxation at large $\kappa$ is explained below in terms of a stiffened effective harmonic potential. In the infinite-$\kappa$ limit, $\langle\Delta\lambda_1\rangle_V$ remains positive but approaches zero. Large $\kappa$ corresponds to disorder realisations that are essentially constant, and thus have no impact on the relaxation rate. The dash-dotted line in \Figref{phasediag}(a) indicates where $\langle\Delta\lambda_1\rangle_V$ reaches its maximum for given $\zeta_T$.

Figure~\figref{phasediag}(b) shows the variance of $\Delta\lambda_1$ as a function of the parameters. The variance is small for small and large $\kappa$, and reaches a maximum at a $\kappa$ of order unity, as indicated by the dash-dotted line.

Other types of Gaussian disorder with a sufficiently smooth correlation function exhibit qualitatively similar behaviours (see \cite{supp}). For exponentially-correlated disorder, by contrast, we find that while the variance behaves similarly to that shown in \Figref{phasediag}(b) for $\zeta_T\lessapprox1$, the mean $\langle \Delta\lambda_1\rangle_V$ is always negative~\cite{supp}. Such non-differentiable disorder generates an infinite number of particle-trapping local minima in $W$, independently of $\kappa$ [see \Figref{pot_reals}(b)]. This provides an intuition for why relaxation is always slower for $C(x)=\exp(-|x|)$.

From our numerical simulations in \Figref{phasediag} we conclude that both the mean and the variance of $\Delta\lambda_1$ have characteristic dependences on the correlation length $\kappa$, and rather simple monotonic dependences on $\zeta_T$. In particular, the characteristic features of the $\kappa$ dependences of both quantities, including their zeros and maxima, remain intact as $\zeta_T$ becomes small, while their magnitudes approach zero. This motivates us to study the problem in the limit of weak disorder.
\begin{figure}
	\includegraphics[width=\linewidth]{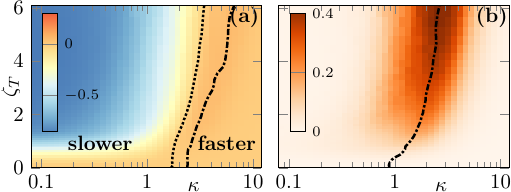}
	\caption{Mean and variance of $\Delta\lambda_1$ from numerical evaluation of \Eqnref{lev} as function of $\kappa$ and $\zeta_T$ using up to $10^4$ Gaussian-correlated disorder realisations. (a) Mean value $\langle\Delta\lambda_1\rangle_V$ (colours). The lines correspond to vanishing (dotted) and maximum (dash-dotted) mean value for given $\zeta_T$. (b) Variance of $\Delta\lambda_1$ (colours). The dash-dotted line shows the location of maximum variance for given $\zeta_T$.}\figlab{phasediag}
\end{figure}
\subsection{Weak-disorder limit}
Using perturbation theory, we first consider again the mean and variance of $\Delta\lambda_1$, but eventually compute the entire distribution of $\Delta\lambda_1$ for $\zeta_T\ll1$. To this end, we split the non-Hermitian operator $\mc{L}^\dagger$ in \Eqnref{lev} into an unperturbed harmonic part and a perturbation. The unperturbed part is brought into Hermitian form using
\algn{
	\ms{H}^\dagger \equiv -\ee^{-x^2/4}\ms{L}^\dagger\ee^{x^2/4}= \ms{H}_0 + \zeta_T\ms{H}^\dagger_V - \frac{1}2\,,
}
where
\algn{\eqnlab{hops}
	\ms{H}_0 = -\partial_x^2 + \frac{x^2}4\,,\quad \ms{H}^\dagger_V = -\kappa^{-1}V'\left(\frac{x}{\kappa}\right)\left(\frac{x}2 +\partial_x\right)\,.
}	
The unperturbed operator $\ms{H}_0$ is Hermitian and given by the Hamiltonian of the quantum harmonic oscillator~\cite{Dav76}, while the transformed perturbation $\ms{H}^\dagger_V$ remains non-Hermitian.
The eigenvalue problem~\eqnref{lev} then reads
\algn{\eqnlab{hermeig}
	\ms{H}^\dagger\Psi_n(x) = \left(\lambda_n+\frac12\right)\Psi_n(x)\,,
}
where $\Psi_n(x) = \mc{N}\ee^{-x^2/4}l_n(x)$ with normalisation $\mc{N}$ chosen so that, using bra-ket notation~\cite{Dav76}, $\langle \Psi_n|\Psi_n\rangle = 1$.

We define raising and lowering operators $a^\dagger$ and $a$ that act on the eigenstates $|n\rangle$, $n=0,1,\ldots$, of $\ms{H}_0=a^\dagger a+1/2$ in the standard way~\cite{Dav76}. In order to express $\ms{H}^\dagger_{V}$ in terms of these ladder operators, we use a Fourier representation to obtain
\algn{\eqnlab{operator}
	\ms{H}_V^\dagger 	= \int_{-\infty}^\infty\!\!\!\ed k\, ik\ee^{-\frac{k^2}2}\kappa\hat V(\kappa k)\ee^{ik a^\dagger}\ee^{ika}a\,,
}
where $\hat V$ denotes the Fourier transform of $V$~\footnote{We use the convention $\hat f(k)=\int_{-\infty}^{\infty}\!\!\ed{x}\ee^{-ikx}f(x)/(2\pi)$ for the Fourier transform of $f$.}.
\subsection{Perturbation theory}
By means of second-order perturbation theory~\cite{Dav76} in $\zeta_T$, we obtain for the relative relaxation rates $\Delta\lambda_n=\lambda_n-n$,
\algn{\eqnlab{secondorder}
	\Delta \lambda_n 	\sim \zeta_T\langle n|\ms{H}_V^\dagger|n\rangle + \zeta_T^2\sum_{m\neq n}\frac{\langle n|\ms{H}_V^\dagger|m\rangle\langle m|\ms{H}_V^\dagger|n\rangle}{n-m}\,.
}

We first focus on the mean and variance of $\Delta\lambda_n$. The rescaled mean $\mu_{\Delta\lambda_n}\equiv \zeta_T^{-2}\langle\Delta\lambda_n\rangle_V$ is obtained by averaging \Eqnref{secondorder} over realisations of $V$, which gives
\sbeqs{\eqnlab{meanvarDl}
\algn{\eqnlab{mean}
	\mu_{\Delta\lambda_n} &\sim \sum_{m\neq n}\frac{\llangle n|\ms{H}_V^\dagger|m\rangle\langle m|\ms{H}_V^\dagger|n\rrangle_V}{n-m}\,,
}
to leading order in $\zeta_T$.

Similarly, the rescaled variance $\sigma^2_{\Delta\lambda_n}\equiv \zeta_T^{-2}(\langle \Delta \lambda_n^2 \rangle_V - \langle\Delta \lambda_n\rangle_V^2)$ of $\Delta\lambda_n$ is obtained as
\algn{\eqnlab{var}
	\sigma^2_{\Delta\lambda_n}\sim\llangle n|\ms{H}_V^\dagger|n\rangle\langle n|\ms{H}_V^\dagger|n\rrangle_V\,.
}
}
Hence, the mean and variance are determined by the disorder-averaged matrix elements $\llangle n|\ms{H}_V^\dagger|m\rangle\langle m|\ms{H}_V^\dagger|n\rrangle_V$, both for $n\neq m$ [\Eqnref{mean}] and for $n=m$ [\Eqnref{var}].
Using \Eqnref{operator} we write these matrix elements as
\algn{
	\llangle n|\ms{H}_V^\dagger|m\rangle\langle m|\ms{H}_V^\dagger|n\rrangle_V = \int_{-\infty}^{\infty}\!\!\!\ed k\,  k^2\ee^{-k^2}\kappa \hat{C}(\kappa k)T_{nm}(k)\,,
}
where $\hat C$ denotes the Fourier transform of $C$ and
\algn{
	T_{nm}(k) 	\equiv& \sqrt{nm}\langle n|\ee^{ik a^\dagger}\ee^{ika}|m-1\rangle\langle m|\ee^{-ik a^\dagger}\ee^{-ika}|n-1\rangle\,,\\
	=&\!\!\!\!\sum_{j_1=0}^{n\wedge(m-1)}\sum_{j_2=0}^{m\wedge(n-1)}\binom{n}{j_1}\binom{m}{j_2}\frac{(k^2)^{n+m-1}(-k^2)^{j_1+j_2}}{(m-1-j_1)!(n-1-j_2)!}\,,\nn
}
with $a\wedge b$ denoting the minimum of $a$ and $b$. Using these expressions, \Eqsref{meanvarDl} are written as
\algn{\eqnlab{meanvarDl2}
	\mu_{\Delta\lambda_n} \sim \int_{-\infty}^{\infty}\!\!\!\ed k\, \hat C(k)K_{n}\left(\frac{k}{\kappa}\right)\,,\ \sigma^2_{\Delta\lambda_n} \sim\int_{-\infty}^{\infty}\!\!\!\ed k\, \hat C(k)M_{n}\left(\frac{k}{\kappa}\right)\,,
}
with
\sbeqs{\eqnlab{kerns}
\algn{
	K_n(k) &\equiv k^2 \ee^{-k^2}\left[\sum_{m=1}^{n} \frac{T_{n(n-m)}(k)}{m} - \sum_{m=1}^{\infty} \frac{T_{n(n+m)}(k)}{m}\right]\,,\\
	M_{n}(k) &\equiv k^2 \ee^{-k^2}T_{nn}(k)\,.
}
}
Equations~\eqnref{meanvarDl2} and \eqnref{kerns} enable us to compute the mean and variance of all relative relaxation rates $\Delta\lambda_n$ at weak disorder.
\subsection{Dominant relaxation rate}
In the long-time limit, relaxation is dominated by the relative relaxation rate $\Delta\lambda_1$, see \Eqnref{lam1}. To determine the statistics of $\Delta\lambda_1$, we need to evaluate \Eqsref{meanvarDl2} for $n=1$. In this case, the integral kernels in \Eqsref{kerns} can be resummed explicitly, yielding
\sbeqs{\eqnlab{kmfuncs}
\algn{
	K_1(k) 	&= k^2\left\{1+\ee^{-k^2}\left[k^2\text{Ein}(-k^2)-1\right]\right\}\,,\\
	M_1(k) 	&= k^4\ee^{-k^2}\,,
}
}
where $\text{Ein(x)}$ denotes the complementary exponential integral~\cite{dlmf}.
\begin{figure}
	\includegraphics[width=\linewidth]{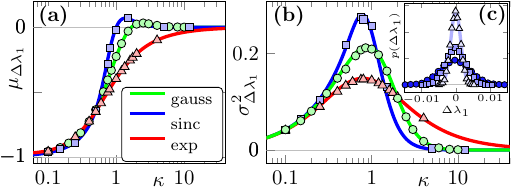}
	\caption{(a) Rescaled mean $\mu_{\Delta\lambda_1}$ as function of $\kappa$ from perturbation theory (solid lines) and numerical simulations (symbols) for different correlation functions. (b) Rescaled variance $\sigma^2_{\Delta\lambda_1}$. (c) Probability distribution $p(\Delta\lambda_1)$ from theory (solid lines) and numerical simulations (symbols) for Gaussian-correlated disorder with $\kappa=0.2$ and different $\zeta_T=0.015$ (bullets), $0.01$ (squares), and $0.005$ (triangles).}\figlab{DE1}
\end{figure}

Figure~\figref{DE1}(a) shows $\mu_{\Delta\lambda_1}$ as function of $\kappa$ for different correlation functions, computed from perturbation theory [\Eqsref{meanvarDl2}] and from small-$\zeta_T$ simulations. We observe that $\mu_{\Delta\lambda_1}$ recovers the non-monotonous dependence on $\kappa$ for Gaussian- and sinc-correlated disorder, observed in \Figref{phasediag}(a). For exponentially correlated disorder, by contrast,  $\mu_{\Delta\lambda_1}$ is monotonous and always negative. Furthermore, we find that $\mu_{\Delta\lambda_1}\to -1$ for small $\kappa$ for all correlation functions, implying that relaxation is slowed down the most by disorder with a small correlation length $\kappa\ll1$. For large $\kappa$, $\mu_{\Delta\lambda_1}$ approaches zero for all correlation functions, just as $\langle\Delta\lambda_1\rangle_V$ in \Figref{phasediag}(a). For the smooth Gaussian and sinc correlation functions, $\mu_{\Delta\lambda_1}$ is positive for large enough $\kappa$ and changes sign at a $\kappa$ of order unity. For intermediate $\kappa$ the behaviour of $\mu_{\Delta\lambda_1}$ is non-universal.

The rescaled variance $\sigma^2_{\Delta\lambda_1}$, shown in \Figref{DE1}(b), exhibits a characteristic maximum at values of $\kappa$ of order unity for all correlation functions, but approaches zero for both small and large $\kappa$, analogous to the behaviour observed in \Figref{phasediag}(b).

Hence, the weak-noise analysis characterised by the rescaled mean $\mu_{\Delta\lambda_1}$ and variance $\sigma^2_{\Delta\lambda_1}$ reproduces the main features observed in \Figref{phasediag}, and it reveals universal limits for these quantities.
\subsection{Universal limits} 
To understand the origins and the generality of the observations made in \Figsref{phasediag} and \figref{DE1} for small and large $\kappa$, we analyse $\mu_{\Delta\lambda_1}$ and $\sigma^2_{\Delta\lambda_1}$ asymptotically in these limits.

Equations~\eqnref{meanvarDl2} show that for small $\kappa$, the expressions for $\mu_{\Delta\lambda_1}$ and $\sigma^2_{\Delta\lambda_1}$ are obtained from large-$k$ expansions of \Eqsref{kmfuncs}, which for $K_1$ gives $K_1(k)\sim -1$. We then obtain
\algn{\eqnlab{Dlamasym1}
	\mu_{\Delta\lambda_1} \sim -1\,,
}
for $\kappa\ll1$, independently of $C$, in agreement with \Figref{DE1}(a).

To interpret \Eqnref{Dlamasym1} physically, we compute the disorder-averaged free energy $F=-\langle \ln Z_\text{eq}\rangle_V$ applying the replica trick $\langle \ln Z_\text{eq}\rangle_V = \dd{m}\langle Z_\text{eq}^m\rangle_V\big |_{m=0}$~\cite{Mez87}. For integer $m$, and using the Gaussian properties of $V$, $\langle Z^m_\text{eq}\rangle_V$ evaluates to
\algn{
	\langle Z^m_\text{eq}\rangle_V=\int_{\mbb{R}^m}\!\!\!\ed^m\ve x\, \exp\left[-\frac{\ve x^2}2 +\frac{\zeta^2_T}{2}\sum_{i,j=1}^m C\left(\frac{x_i-x_j}{\kappa}\right)\right]\,,
}
where the integral is performed over $\ve x \equiv (x_1,\ldots,x_m)^{\sf T}$. Assuming replica symmetry~\cite{Mez87} for $\kappa\ll1$, the non-diagonal terms in the sum vanish and the diagonal terms sum up to $m$, see \Eqnref{corrprop}, independently of $C$. This allows us to evaluate the integral and to use the replica trick. For small $\zeta_T$, the result can be written as $F \sim \ln\left[(1-\zeta_T^2)/2\pi\right]/2$, which agrees with the free energy of a Brownian particle in a harmonic potential with effective stiffness
\algn{\eqnlab{lameff0}
	\alpha^\text{eff}_0 = 1-\zeta_T^2\,.
}
	
By using  $\Delta\lambda_n = (\alpha-1)n$ for harmonic potentials with arbitrary stiffness $\alpha$, we recover the result \eqnref{Dlamasym1} for $\alpha=\alpha^\text{eff}_0$ and $n=1$. This implies that upon introducing a large number of close-by, shallow minima in $W$, which is the case for $\kappa,\zeta_T\ll 1$, the particle experiences a disorder-averaged effective potential. Because of particle trapping, the effective potential is less stiff than the unperturbed one, resulting in a universal correction to $\langle\Delta\lambda_n\rangle_V$ of order $\zeta_T^2$. 

The large-$\kappa$ behaviours of the mean and variance in \Eqsref{meanvarDl2} are obtained from small-$k$ expansions of the integral kernels, leading to 
\algn{\eqnlab{musig}
	\mu_{\Delta\lambda_1} 	\sim \kappa^{-4}C^{(4)}_0\,,\qquad\sigma^2_{\Delta\lambda_1} \sim \kappa^{-4}C^{(4)}_0\,,
}
where $C^{(4)}_0>0$ denotes the fourth derivative of the correlation function evaluated at $x=0$. This shows that relaxation is on average faster for large $\kappa$ whenever $C$ is at least four times differentiable. This is the case for, e.g., the Gaussian and sinc correlation functions in \Figref{DE1}(a).

The mechanism behind the relaxation speedup for $\kappa\gg1$ is understood as follows: For sufficiently smooth $C$ and at large $\kappa$, $W$ is a deformed harmonic potential, see \Figref{pot_reals}(b). Therefore, the relaxation rate is determined by another effective stiffness, $\alpha^\text{eff}_{\infty}$, obtained from the curvature $\alpha^\text{eff}_{\infty}\equiv W''(x_0)$ of $W$ at its minimum $x_0$. To obtain $W''(x_0)$, we expand $V(x/\kappa)$ in a Taylor series in $\kappa^{-1}$ and express $\alpha^\text{eff}_{\infty}$ for $\kappa\gg1$ as
\algn{\eqnlab{lameffinf}
	 \alpha^\text{eff}_{\infty} \sim 1 + \frac{\zeta_T}{\kappa^2}V^{(2)}_0 - \frac{\zeta^2_T}{\kappa^4}V^{(1)}_0V^{(3)}_0\,,
}
where we denote as $V_0^{(n)}$ the $n^\text{th}$ derivative of $V$ at $x=0$. Using again $\Delta\lambda_n = (\alpha-1)n$, but now with $\alpha=\alpha^\text{eff}_\infty$ and $n=1$, we recover \Eqsref{musig} by computing the rescaled mean and average. Equation~\eqnref{lameffinf} implies an effective average stiffening $\langle \alpha^\text{eff}_{\infty}\rangle_V-1>0$ of the harmonic potential at large $\kappa$, which originates from the anticorrelation $\langle V_0^{(1)}V_0^{(3)}\rangle_V=-C_0^{(4)}<0$, a general property of smooth Gaussian random fields.

The exponential correlation function, by contrast, is not differentiable at $x=0$. Consequently, $\mu_{\Delta\lambda_1}$ need not be positive for large $\kappa$, in agreement with \Figref{DE1}(a).

Taken together, for weak Gaussian disorder with sufficiently smooth correlation function, the Brownian particle experiences effective harmonic potentials that are either less stiff [small $\kappa$, \Eqnref{lameff0}] or on average stiffer [large $\kappa$, \Eqnref{lameffinf}] than in the unperturbed case. Hence, there exists a finite $\kappa$ for which $\langle\Delta\lambda_1\rangle_V$ changes sign. As a consequence, a suitable choice of $\kappa$ exponentially increases or decreases the speed of relaxation, whenever the correlation function is sufficiently smooth.
\subsection{Gaussian statistics} 
We now show that for $\zeta_T\ll1$, the statistics of $\Delta\lambda_1$ is Gaussian, and thus completely determined by the mean and variance computed above. This is not true in general, since the eigenvalues are highly non-linear functions of the disorder fields $V$~\cite{Ben78}, in particular for non-Hermitian operators such as $\ms{L}^\dagger$~\cite{Tre05}.

To obtain the statistics of $\Delta\lambda_1$, we compute the centered $n^\text{th}$ moments of $\Delta\lambda_1$ to leading order in $\zeta_T$. These moments behave as $\sim (n-1)!!\zeta_T^{n}\sigma^n_{\Delta\lambda_1}$ for even $n$, where $n!!$ denotes the double factorial, and $\sim0$ for odd $n$. We infer that the probability distribution $p(\Delta\lambda_1)$ is Gaussian with mean $\zeta_T^2\mu_{\Delta\lambda_1}$ and variance $\zeta_T^2\sigma^2_{\Delta\lambda_1}$ for $\zeta_T\ll1$. In the limit $\zeta_T\to0$, $p(\Delta\lambda_1)$ converges to a delta-function centered at $\Delta\lambda_1=0$, as expected.

Figure~\figref{DE1}(c) shows $p(\Delta\lambda_1)$ for given $\kappa$ and different values of $\zeta_T$ from theory and simulations. In agreement between both methods, $p(\Delta\lambda_1)$ has a Gaussian shape that focuses around $\Delta\lambda_1=0$ as $\zeta_T$ decreases.
\section{Conclusions}
For a simple model of complex energy landscapes, we have shown that smooth correlated disorder may both decrease and increase the asymptotic relaxation rate. This is explained by particle trapping in local minima (small $\kappa$) and an effectively stiffened harmonic potential (large $\kappa$). The statistics of the relaxation rate is Gaussian at weak disorder.

Our results imply that correlated disorder strongly impacts relaxation, even in simple systems such as the one considered here. This seems to have not been recognised before, although the effect is stronger than that associated with other relaxation anomalies~\cite{Lu17,Lap20}.

Our predictions are directly testable in experiments involving, e.g., colloids in laser traps superimposed with disorder in the form of speckles, whose amplitude and correlation length can be tuned by modifying the speckle intensity and the beam size~\cite{Han12,Bew16a}. An important open question is whether comparable disorder-induced effects persist in more complex systems. Promising next steps are to analyse the problem in dimensions higher than one, thus introducing additional pathways to escape potential minima.
\begin{acknowledgments}
\end{acknowledgments}

\begin{thebibliography}{65}%
\makeatletter
\providecommand \@ifxundefined [1]{%
 \@ifx{#1\undefined}
}%
\providecommand \@ifnum [1]{%
 \ifnum #1\expandafter \@firstoftwo
 \else \expandafter \@secondoftwo
 \fi
}%
\providecommand \@ifx [1]{%
 \ifx #1\expandafter \@firstoftwo
 \else \expandafter \@secondoftwo
 \fi
}%
\providecommand \natexlab [1]{#1}%
\providecommand \enquote  [1]{``#1''}%
\providecommand \bibnamefont  [1]{#1}%
\providecommand \bibfnamefont [1]{#1}%
\providecommand \citenamefont [1]{#1}%
\providecommand \href@noop [0]{\@secondoftwo}%
\providecommand \href [0]{\begingroup \@sanitize@url \@href}%
\providecommand \@href[1]{\@@startlink{#1}\@@href}%
\providecommand \@@href[1]{\endgroup#1\@@endlink}%
\providecommand \@sanitize@url [0]{\catcode `\\12\catcode `\$12\catcode
  `\&12\catcode `\#12\catcode `\^12\catcode `\_12\catcode `\%12\relax}%
\providecommand \@@startlink[1]{}%
\providecommand \@@endlink[0]{}%
\providecommand \url  [0]{\begingroup\@sanitize@url \@url }%
\providecommand \@url [1]{\endgroup\@href {#1}{\urlprefix }}%
\providecommand \urlprefix  [0]{URL }%
\providecommand \Eprint [0]{\href }%
\providecommand \doibase [0]{https://doi.org/}%
\providecommand \selectlanguage [0]{\@gobble}%
\providecommand \bibinfo  [0]{\@secondoftwo}%
\providecommand \bibfield  [0]{\@secondoftwo}%
\providecommand \translation [1]{[#1]}%
\providecommand \BibitemOpen [0]{}%
\providecommand \bibitemStop [0]{}%
\providecommand \bibitemNoStop [0]{.\EOS\space}%
\providecommand \EOS [0]{\spacefactor3000\relax}%
\providecommand \BibitemShut  [1]{\csname bibitem#1\endcsname}%
\let\auto@bib@innerbib\@empty
\bibitem [{\citenamefont {Mpemba}\ and\ \citenamefont {Osborne}(1969)}]{Mpe69}%
  \BibitemOpen
  \bibfield  {author} {\bibinfo {author} {\bibfnamefont {E.~B.}\ \bibnamefont
  {Mpemba}}\ and\ \bibinfo {author} {\bibfnamefont {D.~G.}\ \bibnamefont
  {Osborne}},\ }\bibfield  {title} {\bibinfo {title} {{Cool?}},\ }\href@noop {}
  {\bibfield  {journal} {\bibinfo  {journal} {Physics Education}\ }\textbf
  {\bibinfo {volume} {4}},\ \bibinfo {pages} {172} (\bibinfo {year}
  {1969})}\BibitemShut {NoStop}%
\bibitem [{\citenamefont {Baity-Jesi}\ \emph {et~al.}(2019)\citenamefont
  {Baity-Jesi}, \citenamefont {Calore}, \citenamefont {Cruz}, \citenamefont
  {Fernandez}, \citenamefont {Gil-Narvi{\'{o}}n}, \citenamefont
  {Gordillo-Guerrero}, \citenamefont {I{\~{n}}iguez}, \citenamefont {Lasanta},
  \citenamefont {Maiorano}, \citenamefont {Marinari},\ and\ \citenamefont
  {Others}}]{Bai19}%
  \BibitemOpen
  \bibfield  {author} {\bibinfo {author} {\bibfnamefont {M.}~\bibnamefont
  {Baity-Jesi}}, \bibinfo {author} {\bibfnamefont {E.}~\bibnamefont {Calore}},
  \bibinfo {author} {\bibfnamefont {A.}~\bibnamefont {Cruz}}, \bibinfo {author}
  {\bibfnamefont {L.~A.}\ \bibnamefont {Fernandez}}, \bibinfo {author}
  {\bibfnamefont {J.~M.}\ \bibnamefont {Gil-Narvi{\'{o}}n}}, \bibinfo {author}
  {\bibfnamefont {A.}~\bibnamefont {Gordillo-Guerrero}}, \bibinfo {author}
  {\bibfnamefont {D.}~\bibnamefont {I{\~{n}}iguez}}, \bibinfo {author}
  {\bibfnamefont {A.}~\bibnamefont {Lasanta}}, \bibinfo {author} {\bibfnamefont
  {A.}~\bibnamefont {Maiorano}}, \bibinfo {author} {\bibfnamefont
  {E.}~\bibnamefont {Marinari}},\ and\ \bibinfo {author} {\bibnamefont
  {Others}},\ }\bibfield  {title} {\bibinfo {title} {{The Mpemba effect in spin
  glasses is a persistent memory effect}},\ }\href@noop {} {\bibfield
  {journal} {\bibinfo  {journal} {Proceedings of the National Academy of
  Sciences}\ }\textbf {\bibinfo {volume} {116}},\ \bibinfo {pages} {15350}
  (\bibinfo {year} {2019})}\BibitemShut {NoStop}%
\bibitem [{\citenamefont {Arceri}\ \emph {et~al.}(2022)\citenamefont {Arceri},
  \citenamefont {Landes}, \citenamefont {Berthier},\ and\ \citenamefont
  {Biroli}}]{Arc22}%
  \BibitemOpen
  \bibfield  {author} {\bibinfo {author} {\bibfnamefont {F.}~\bibnamefont
  {Arceri}}, \bibinfo {author} {\bibfnamefont {F.~P.}\ \bibnamefont {Landes}},
  \bibinfo {author} {\bibfnamefont {L.}~\bibnamefont {Berthier}},\ and\
  \bibinfo {author} {\bibfnamefont {G.}~\bibnamefont {Biroli}},\ }\bibfield
  {title} {\bibinfo {title} {{Glasses and aging, A statistical mechanics
  perspective}},\ }in\ \href@noop {} {\emph {\bibinfo {booktitle} {Statistical
  and Nonlinear Physics}}}\ (\bibinfo  {publisher} {Springer},\ \bibinfo {year}
  {2022})\ pp.\ \bibinfo {pages} {229--296}\BibitemShut {NoStop}%
\bibitem [{\citenamefont {Kovacs}\ \emph {et~al.}(1963)\citenamefont {Kovacs},
  \citenamefont {Stratton},\ and\ \citenamefont {Ferry}}]{Kov63}%
  \BibitemOpen
  \bibfield  {author} {\bibinfo {author} {\bibfnamefont {A.~J.}\ \bibnamefont
  {Kovacs}}, \bibinfo {author} {\bibfnamefont {R.~A.}\ \bibnamefont
  {Stratton}},\ and\ \bibinfo {author} {\bibfnamefont {J.~D.}\ \bibnamefont
  {Ferry}},\ }\bibfield  {title} {\bibinfo {title} {{Dynamic mechanical
  properties of polyvinyl acetate in shear in the glass transition temperature
  range}},\ }\href@noop {} {\bibfield  {journal} {\bibinfo  {journal} {The
  Journal of Physical Chemistry}\ }\textbf {\bibinfo {volume} {67}},\ \bibinfo
  {pages} {152} (\bibinfo {year} {1963})}\BibitemShut {NoStop}%
\bibitem [{\citenamefont {Calabrese}\ \emph {et~al.}(2006)\citenamefont
  {Calabrese}, \citenamefont {Gambassi},\ and\ \citenamefont
  {Krzakala}}]{Cal06}%
  \BibitemOpen
  \bibfield  {author} {\bibinfo {author} {\bibfnamefont {P.}~\bibnamefont
  {Calabrese}}, \bibinfo {author} {\bibfnamefont {A.}~\bibnamefont
  {Gambassi}},\ and\ \bibinfo {author} {\bibfnamefont {F.}~\bibnamefont
  {Krzakala}},\ }\bibfield  {title} {\bibinfo {title} {{Critical ageing of
  Ising ferromagnets relaxing from an ordered state}},\ }\href@noop {}
  {\bibfield  {journal} {\bibinfo  {journal} {Journal of Statistical Mechanics:
  Theory and Experiment}\ }\textbf {\bibinfo {volume} {2006}},\ \bibinfo
  {pages} {P06016} (\bibinfo {year} {2006})}\BibitemShut {NoStop}%
\bibitem [{\citenamefont {Bray}(2002)}]{Bra02}%
  \BibitemOpen
  \bibfield  {author} {\bibinfo {author} {\bibfnamefont {A.~J.}\ \bibnamefont
  {Bray}},\ }\bibfield  {title} {\bibinfo {title} {{Theory of phase-ordering
  kinetics}},\ }\href@noop {} {\bibfield  {journal} {\bibinfo  {journal}
  {Advances in Physics}\ }\textbf {\bibinfo {volume} {51}},\ \bibinfo {pages}
  {481} (\bibinfo {year} {2002})}\BibitemShut {NoStop}%
\bibitem [{\citenamefont {Vadakkayil}\ \emph {et~al.}(2024)\citenamefont
  {Vadakkayil}, \citenamefont {Esposito},\ and\ \citenamefont
  {Meibohm}}]{Vad24}%
  \BibitemOpen
  \bibfield  {author} {\bibinfo {author} {\bibfnamefont {N.}~\bibnamefont
  {Vadakkayil}}, \bibinfo {author} {\bibfnamefont {M.}~\bibnamefont
  {Esposito}},\ and\ \bibinfo {author} {\bibfnamefont {J.}~\bibnamefont
  {Meibohm}},\ }\bibfield  {title} {\bibinfo {title} {{Critical fluctuations at
  a finite-time dynamical phase transition}},\ }\href@noop {} {\bibfield
  {journal} {\bibinfo  {journal} {Physical Review E}\ }\textbf {\bibinfo
  {volume} {110}},\ \bibinfo {pages} {064156} (\bibinfo {year}
  {2024})}\BibitemShut {NoStop}%
\bibitem [{\citenamefont {Lu}\ and\ \citenamefont {Raz}(2017)}]{Lu17}%
  \BibitemOpen
  \bibfield  {author} {\bibinfo {author} {\bibfnamefont {Z.}~\bibnamefont
  {Lu}}\ and\ \bibinfo {author} {\bibfnamefont {O.}~\bibnamefont {Raz}},\
  }\bibfield  {title} {\bibinfo {title} {{Nonequilibrium thermodynamics of the
  Markovian Mpemba effect and its inverse}},\ }\href@noop {} {\bibfield
  {journal} {\bibinfo  {journal} {Proceedings of the National Academy of
  Sciences}\ }\textbf {\bibinfo {volume} {114}},\ \bibinfo {pages} {5083}
  (\bibinfo {year} {2017})}\BibitemShut {NoStop}%
\bibitem [{\citenamefont {Klich}\ \emph {et~al.}(2019)\citenamefont {Klich},
  \citenamefont {Raz}, \citenamefont {Hirschberg},\ and\ \citenamefont
  {Vucelja}}]{Kli19}%
  \BibitemOpen
  \bibfield  {author} {\bibinfo {author} {\bibfnamefont {I.}~\bibnamefont
  {Klich}}, \bibinfo {author} {\bibfnamefont {O.}~\bibnamefont {Raz}}, \bibinfo
  {author} {\bibfnamefont {O.}~\bibnamefont {Hirschberg}},\ and\ \bibinfo
  {author} {\bibfnamefont {M.}~\bibnamefont {Vucelja}},\ }\bibfield  {title}
  {\bibinfo {title} {{Mpemba index and anomalous relaxation}},\ }\href@noop {}
  {\bibfield  {journal} {\bibinfo  {journal} {Physical Review X}\ }\textbf
  {\bibinfo {volume} {9}},\ \bibinfo {pages} {021060} (\bibinfo {year}
  {2019})}\BibitemShut {NoStop}%
\bibitem [{\citenamefont {Kumar}\ and\ \citenamefont
  {Bechhoefer}(2020)}]{Kum20}%
  \BibitemOpen
  \bibfield  {author} {\bibinfo {author} {\bibfnamefont {A.}~\bibnamefont
  {Kumar}}\ and\ \bibinfo {author} {\bibfnamefont {J.}~\bibnamefont
  {Bechhoefer}},\ }\bibfield  {title} {\bibinfo {title} {{Exponentially faster
  cooling in a colloidal system}},\ }\href@noop {} {\bibfield  {journal}
  {\bibinfo  {journal} {Nature}\ }\textbf {\bibinfo {volume} {584}},\ \bibinfo
  {pages} {64} (\bibinfo {year} {2020})}\BibitemShut {NoStop}%
\bibitem [{\citenamefont {Ch{\'{e}}trite}\ \emph {et~al.}(2021)\citenamefont
  {Ch{\'{e}}trite}, \citenamefont {Kumar},\ and\ \citenamefont
  {Bechhoefer}}]{Che21}%
  \BibitemOpen
  \bibfield  {author} {\bibinfo {author} {\bibfnamefont {R.}~\bibnamefont
  {Ch{\'{e}}trite}}, \bibinfo {author} {\bibfnamefont {A.}~\bibnamefont
  {Kumar}},\ and\ \bibinfo {author} {\bibfnamefont {J.}~\bibnamefont
  {Bechhoefer}},\ }\bibfield  {title} {\bibinfo {title} {{The metastable Mpemba
  effect corresponds to a non-monotonic temperature dependence of extractable
  work}},\ }\href@noop {} {\bibfield  {journal} {\bibinfo  {journal} {Frontiers
  in Physics}\ }\textbf {\bibinfo {volume} {9}},\ \bibinfo {pages} {141}
  (\bibinfo {year} {2021})}\BibitemShut {NoStop}%
\bibitem [{\citenamefont {Lapolla}\ and\ \citenamefont {Godec}(2020)}]{Lap20}%
  \BibitemOpen
  \bibfield  {author} {\bibinfo {author} {\bibfnamefont {A.}~\bibnamefont
  {Lapolla}}\ and\ \bibinfo {author} {\bibfnamefont {A.}~\bibnamefont
  {Godec}},\ }\bibfield  {title} {\bibinfo {title} {{Faster uphill relaxation
  in thermodynamically equidistant temperature quenches}},\ }\href@noop {}
  {\bibfield  {journal} {\bibinfo  {journal} {Physical Review Letters}\
  }\textbf {\bibinfo {volume} {125}},\ \bibinfo {pages} {110602} (\bibinfo
  {year} {2020})}\BibitemShut {NoStop}%
\bibitem [{\citenamefont {Meibohm}\ \emph {et~al.}(2021)\citenamefont
  {Meibohm}, \citenamefont {Forastiere}, \citenamefont {Adeleke-Larodo},\ and\
  \citenamefont {Proesmans}}]{Mei21b}%
  \BibitemOpen
  \bibfield  {author} {\bibinfo {author} {\bibfnamefont {J.}~\bibnamefont
  {Meibohm}}, \bibinfo {author} {\bibfnamefont {D.}~\bibnamefont {Forastiere}},
  \bibinfo {author} {\bibfnamefont {T.}~\bibnamefont {Adeleke-Larodo}},\ and\
  \bibinfo {author} {\bibfnamefont {K.}~\bibnamefont {Proesmans}},\ }\bibfield
  {title} {\bibinfo {title} {{Relaxation-speed crossover in anharmonic
  potentials}},\ }\href@noop {} {\bibfield  {journal} {\bibinfo  {journal}
  {Physical Review E}\ }\textbf {\bibinfo {volume} {104}},\ \bibinfo {pages}
  {L032105} (\bibinfo {year} {2021})}\BibitemShut {NoStop}%
\bibitem [{\citenamefont {{Van Vu}}\ and\ \citenamefont
  {Hasegawa}(2021)}]{Vu21}%
  \BibitemOpen
  \bibfield  {author} {\bibinfo {author} {\bibfnamefont {T.}~\bibnamefont {{Van
  Vu}}}\ and\ \bibinfo {author} {\bibfnamefont {Y.}~\bibnamefont {Hasegawa}},\
  }\bibfield  {title} {\bibinfo {title} {{Toward relaxation asymmetry: Heating
  is faster than cooling}},\ }\href@noop {} {\bibfield  {journal} {\bibinfo
  {journal} {Physical Review Research}\ }\textbf {\bibinfo {volume} {3}},\
  \bibinfo {pages} {043160} (\bibinfo {year} {2021})}\BibitemShut {NoStop}%
\bibitem [{\citenamefont {Iba{\~{n}}ez}\ \emph {et~al.}(2024)\citenamefont
  {Iba{\~{n}}ez}, \citenamefont {Dieball}, \citenamefont {Lasanta},
  \citenamefont {Godec},\ and\ \citenamefont {Rica}}]{Iba24}%
  \BibitemOpen
  \bibfield  {author} {\bibinfo {author} {\bibfnamefont {M.}~\bibnamefont
  {Iba{\~{n}}ez}}, \bibinfo {author} {\bibfnamefont {K.}~\bibnamefont
  {Dieball}}, \bibinfo {author} {\bibfnamefont {A.}~\bibnamefont {Lasanta}},
  \bibinfo {author} {\bibfnamefont {A.}~\bibnamefont {Godec}},\ and\ \bibinfo
  {author} {\bibfnamefont {R.}~\bibnamefont {Rica}},\ }\bibfield  {title}
  {\bibinfo {title} {{Heating and cooling are fundamentally asymmetric and
  evolve along distinct pathways}},\ }\href@noop {} {\bibfield  {journal}
  {\bibinfo  {journal} {Nature Physics}\ }\textbf {\bibinfo {volume} {20}},\
  \bibinfo {pages} {1} (\bibinfo {year} {2024})}\BibitemShut {NoStop}%
\bibitem [{\citenamefont {Meibohm}\ and\ \citenamefont
  {Esposito}(2022)}]{Mei22a}%
  \BibitemOpen
  \bibfield  {author} {\bibinfo {author} {\bibfnamefont {J.}~\bibnamefont
  {Meibohm}}\ and\ \bibinfo {author} {\bibfnamefont {M.}~\bibnamefont
  {Esposito}},\ }\bibfield  {title} {\bibinfo {title} {{Finite-Time Dynamical
  Phase Transition in Nonequilibrium Relaxation}},\ }\href@noop {} {\bibfield
  {journal} {\bibinfo  {journal} {Physical Review Letters}\ }\textbf {\bibinfo
  {volume} {128}},\ \bibinfo {pages} {110603} (\bibinfo {year}
  {2022})}\BibitemShut {NoStop}%
\bibitem [{\citenamefont {Meibohm}\ and\ \citenamefont
  {Esposito}(2023)}]{Mei23b}%
  \BibitemOpen
  \bibfield  {author} {\bibinfo {author} {\bibfnamefont {J.}~\bibnamefont
  {Meibohm}}\ and\ \bibinfo {author} {\bibfnamefont {M.}~\bibnamefont
  {Esposito}},\ }\bibfield  {title} {\bibinfo {title} {{Landau theory for
  finite-time dynamical phase transitions}},\ }\href@noop {} {\bibfield
  {journal} {\bibinfo  {journal} {New Journal of Physics}\ }\textbf {\bibinfo
  {volume} {25}},\ \bibinfo {pages} {023034} (\bibinfo {year}
  {2023})}\BibitemShut {NoStop}%
\bibitem [{\citenamefont {Teza}\ \emph
  {et~al.}(2023{\natexlab{a}})\citenamefont {Teza}, \citenamefont {Yaacoby},\
  and\ \citenamefont {Raz}}]{Tez23b}%
  \BibitemOpen
  \bibfield  {author} {\bibinfo {author} {\bibfnamefont {G.}~\bibnamefont
  {Teza}}, \bibinfo {author} {\bibfnamefont {R.}~\bibnamefont {Yaacoby}},\ and\
  \bibinfo {author} {\bibfnamefont {O.}~\bibnamefont {Raz}},\ }\bibfield
  {title} {\bibinfo {title} {{Eigenvalue crossing as a phase transition in
  relaxation dynamics}},\ }\href@noop {} {\bibfield  {journal} {\bibinfo
  {journal} {Physical Review Letters}\ }\textbf {\bibinfo {volume} {130}},\
  \bibinfo {pages} {207103} (\bibinfo {year} {2023}{\natexlab{a}})}\BibitemShut
  {NoStop}%
\bibitem [{\citenamefont {Chatterjee}\ \emph {et~al.}(2023)\citenamefont
  {Chatterjee}, \citenamefont {Takada},\ and\ \citenamefont
  {Hayakawa}}]{Cha23}%
  \BibitemOpen
  \bibfield  {author} {\bibinfo {author} {\bibfnamefont {A.~K.}\ \bibnamefont
  {Chatterjee}}, \bibinfo {author} {\bibfnamefont {S.}~\bibnamefont {Takada}},\
  and\ \bibinfo {author} {\bibfnamefont {H.}~\bibnamefont {Hayakawa}},\
  }\bibfield  {title} {\bibinfo {title} {{Quantum Mpemba effect in a quantum
  dot with reservoirs}},\ }\href@noop {} {\bibfield  {journal} {\bibinfo
  {journal} {Physical Review Letters}\ }\textbf {\bibinfo {volume} {131}},\
  \bibinfo {pages} {080402} (\bibinfo {year} {2023})}\BibitemShut {NoStop}%
\bibitem [{\citenamefont {Carollo}\ \emph {et~al.}(2021)\citenamefont
  {Carollo}, \citenamefont {Lasanta},\ and\ \citenamefont
  {Lesanovsky}}]{Car21}%
  \BibitemOpen
  \bibfield  {author} {\bibinfo {author} {\bibfnamefont {F.}~\bibnamefont
  {Carollo}}, \bibinfo {author} {\bibfnamefont {A.}~\bibnamefont {Lasanta}},\
  and\ \bibinfo {author} {\bibfnamefont {I.}~\bibnamefont {Lesanovsky}},\
  }\bibfield  {title} {\bibinfo {title} {{Exponentially accelerated approach to
  stationarity in Markovian open quantum systems through the Mpemba effect}},\
  }\href@noop {} {\bibfield  {journal} {\bibinfo  {journal} {Physical Review
  Letters}\ }\textbf {\bibinfo {volume} {127}},\ \bibinfo {pages} {060401}
  (\bibinfo {year} {2021})}\BibitemShut {NoStop}%
\bibitem [{\citenamefont {Wu}\ \emph {et~al.}(2024)\citenamefont {Wu},
  \citenamefont {Nettersheim}, \citenamefont {Fe{\ss}}, \citenamefont
  {Schnell}, \citenamefont {Burgardt}, \citenamefont {Hiebel}, \citenamefont
  {Adam}, \citenamefont {Eckardt},\ and\ \citenamefont {Widera}}]{Wu24}%
  \BibitemOpen
  \bibfield  {author} {\bibinfo {author} {\bibfnamefont {L.-N.}\ \bibnamefont
  {Wu}}, \bibinfo {author} {\bibfnamefont {J.}~\bibnamefont {Nettersheim}},
  \bibinfo {author} {\bibfnamefont {J.}~\bibnamefont {Fe{\ss}}}, \bibinfo
  {author} {\bibfnamefont {A.}~\bibnamefont {Schnell}}, \bibinfo {author}
  {\bibfnamefont {S.}~\bibnamefont {Burgardt}}, \bibinfo {author}
  {\bibfnamefont {S.}~\bibnamefont {Hiebel}}, \bibinfo {author} {\bibfnamefont
  {D.}~\bibnamefont {Adam}}, \bibinfo {author} {\bibfnamefont {A.}~\bibnamefont
  {Eckardt}},\ and\ \bibinfo {author} {\bibfnamefont {A.}~\bibnamefont
  {Widera}},\ }\bibfield  {title} {\bibinfo {title} {{Indication of critical
  scaling in time during the relaxation of an open quantum system}},\
  }\href@noop {} {\bibfield  {journal} {\bibinfo  {journal} {Nature
  Communications}\ }\textbf {\bibinfo {volume} {15}},\ \bibinfo {pages} {1714}
  (\bibinfo {year} {2024})}\BibitemShut {NoStop}%
\bibitem [{\citenamefont {Joshi}\ \emph {et~al.}(2024)\citenamefont {Joshi},
  \citenamefont {Franke}, \citenamefont {Rath}, \citenamefont {Ares},
  \citenamefont {Murciano}, \citenamefont {Kranzl}, \citenamefont {Blatt},
  \citenamefont {Zoller}, \citenamefont {Vermersch}, \citenamefont
  {Calabrese},\ and\ \citenamefont {Others}}]{Jos24}%
  \BibitemOpen
  \bibfield  {author} {\bibinfo {author} {\bibfnamefont {L.~K.}\ \bibnamefont
  {Joshi}}, \bibinfo {author} {\bibfnamefont {J.}~\bibnamefont {Franke}},
  \bibinfo {author} {\bibfnamefont {A.}~\bibnamefont {Rath}}, \bibinfo {author}
  {\bibfnamefont {F.}~\bibnamefont {Ares}}, \bibinfo {author} {\bibfnamefont
  {S.}~\bibnamefont {Murciano}}, \bibinfo {author} {\bibfnamefont
  {F.}~\bibnamefont {Kranzl}}, \bibinfo {author} {\bibfnamefont
  {R.}~\bibnamefont {Blatt}}, \bibinfo {author} {\bibfnamefont
  {P.}~\bibnamefont {Zoller}}, \bibinfo {author} {\bibfnamefont
  {B.}~\bibnamefont {Vermersch}}, \bibinfo {author} {\bibfnamefont
  {P.}~\bibnamefont {Calabrese}},\ and\ \bibinfo {author} {\bibnamefont
  {Others}},\ }\bibfield  {title} {\bibinfo {title} {{Observing the quantum
  Mpemba effect in quantum simulations}},\ }\href@noop {} {\bibfield  {journal}
  {\bibinfo  {journal} {Physical Review Letters}\ }\textbf {\bibinfo {volume}
  {133}},\ \bibinfo {pages} {010402} (\bibinfo {year} {2024})}\BibitemShut
  {NoStop}%
\bibitem [{\citenamefont {Rylands}\ \emph {et~al.}(2024)\citenamefont
  {Rylands}, \citenamefont {Klobas}, \citenamefont {Ares}, \citenamefont
  {Calabrese}, \citenamefont {Murciano},\ and\ \citenamefont
  {Bertini}}]{Ryl24}%
  \BibitemOpen
  \bibfield  {author} {\bibinfo {author} {\bibfnamefont {C.}~\bibnamefont
  {Rylands}}, \bibinfo {author} {\bibfnamefont {K.}~\bibnamefont {Klobas}},
  \bibinfo {author} {\bibfnamefont {F.}~\bibnamefont {Ares}}, \bibinfo {author}
  {\bibfnamefont {P.}~\bibnamefont {Calabrese}}, \bibinfo {author}
  {\bibfnamefont {S.}~\bibnamefont {Murciano}},\ and\ \bibinfo {author}
  {\bibfnamefont {B.}~\bibnamefont {Bertini}},\ }\bibfield  {title} {\bibinfo
  {title} {{Microscopic origin of the quantum Mpemba effect in integrable
  systems}},\ }\href@noop {} {\bibfield  {journal} {\bibinfo  {journal}
  {Physical Review Letters}\ }\textbf {\bibinfo {volume} {133}},\ \bibinfo
  {pages} {010401} (\bibinfo {year} {2024})}\BibitemShut {NoStop}%
\bibitem [{\citenamefont {Ptaszynski}\ and\ \citenamefont
  {Esposito}(2025)}]{Pta24}%
  \BibitemOpen
  \bibfield  {author} {\bibinfo {author} {\bibfnamefont {K.}~\bibnamefont
  {Ptaszynski}}\ and\ \bibinfo {author} {\bibfnamefont {M.}~\bibnamefont
  {Esposito}},\ }\bibfield  {title} {\bibinfo {title} {{Open-system eigenstate
  thermalization in a noninteracting integrable model}},\ }\href@noop {}
  {\bibfield  {journal} {\bibinfo  {journal} {Physical Review E}\ }\textbf
  {\bibinfo {volume} {111}},\ \bibinfo {pages} {014129} (\bibinfo {year}
  {2025})}\BibitemShut {NoStop}%
\bibitem [{\citenamefont {Deffner}\ \emph {et~al.}(2014)\citenamefont
  {Deffner}, \citenamefont {Jarzynski},\ and\ \citenamefont {del
  Campo}}]{Def14}%
  \BibitemOpen
  \bibfield  {author} {\bibinfo {author} {\bibfnamefont {S.}~\bibnamefont
  {Deffner}}, \bibinfo {author} {\bibfnamefont {C.}~\bibnamefont {Jarzynski}},\
  and\ \bibinfo {author} {\bibfnamefont {A.}~\bibnamefont {del Campo}},\
  }\bibfield  {title} {\bibinfo {title} {{Classical and quantum shortcuts to
  adiabaticity for scale-invariant driving}},\ }\href@noop {} {\bibfield
  {journal} {\bibinfo  {journal} {Physical Review X}\ }\textbf {\bibinfo
  {volume} {4}},\ \bibinfo {pages} {021013} (\bibinfo {year}
  {2014})}\BibitemShut {NoStop}%
\bibitem [{\citenamefont {Teza}\ \emph
  {et~al.}(2023{\natexlab{b}})\citenamefont {Teza}, \citenamefont {Yaacoby},\
  and\ \citenamefont {Raz}}]{Tez23a}%
  \BibitemOpen
  \bibfield  {author} {\bibinfo {author} {\bibfnamefont {G.}~\bibnamefont
  {Teza}}, \bibinfo {author} {\bibfnamefont {R.}~\bibnamefont {Yaacoby}},\ and\
  \bibinfo {author} {\bibfnamefont {O.}~\bibnamefont {Raz}},\ }\bibfield
  {title} {\bibinfo {title} {{Relaxation shortcuts through boundary
  coupling}},\ }\href@noop {} {\bibfield  {journal} {\bibinfo  {journal}
  {Physical Review Letters}\ }\textbf {\bibinfo {volume} {131}},\ \bibinfo
  {pages} {017101} (\bibinfo {year} {2023}{\natexlab{b}})}\BibitemShut
  {NoStop}%
\bibitem [{\citenamefont {Gu{\'{e}}ry-Odelin}\ \emph
  {et~al.}(2023)\citenamefont {Gu{\'{e}}ry-Odelin}, \citenamefont {Jarzynski},
  \citenamefont {Plata}, \citenamefont {Prados},\ and\ \citenamefont
  {Trizac}}]{Gue23}%
  \BibitemOpen
  \bibfield  {author} {\bibinfo {author} {\bibfnamefont {D.}~\bibnamefont
  {Gu{\'{e}}ry-Odelin}}, \bibinfo {author} {\bibfnamefont {C.}~\bibnamefont
  {Jarzynski}}, \bibinfo {author} {\bibfnamefont {C.~A.}\ \bibnamefont
  {Plata}}, \bibinfo {author} {\bibfnamefont {A.}~\bibnamefont {Prados}},\ and\
  \bibinfo {author} {\bibfnamefont {E.}~\bibnamefont {Trizac}},\ }\bibfield
  {title} {\bibinfo {title} {{Driving rapidly while remaining in control:
  classical shortcuts from Hamiltonian to stochastic dynamics}},\ }\href@noop
  {} {\bibfield  {journal} {\bibinfo  {journal} {Reports on Progress in
  Physics}\ }\textbf {\bibinfo {volume} {86}},\ \bibinfo {pages} {35902}
  (\bibinfo {year} {2023})}\BibitemShut {NoStop}%
\bibitem [{\citenamefont {Gal}\ and\ \citenamefont {Raz}(2020)}]{Gal20}%
  \BibitemOpen
  \bibfield  {author} {\bibinfo {author} {\bibfnamefont {A.}~\bibnamefont
  {Gal}}\ and\ \bibinfo {author} {\bibfnamefont {O.}~\bibnamefont {Raz}},\
  }\bibfield  {title} {\bibinfo {title} {{Precooling strategy allows
  exponentially faster heating}},\ }\href@noop {} {\bibfield  {journal}
  {\bibinfo  {journal} {Physical Review Letters}\ }\textbf {\bibinfo {volume}
  {124}},\ \bibinfo {pages} {060602} (\bibinfo {year} {2020})}\BibitemShut
  {NoStop}%
\bibitem [{\citenamefont {{Gonz{\'{a}}lez-Adalid Pemart\'in}}\ \emph
  {et~al.}(2021)\citenamefont {{Gonz{\'{a}}lez-Adalid Pemart\'in}},
  \citenamefont {Momp{\'{o}}}, \citenamefont {Lasanta}, \citenamefont
  {Mart\'in-Mayor},\ and\ \citenamefont {Salas}}]{Pem21}%
  \BibitemOpen
  \bibfield  {author} {\bibinfo {author} {\bibfnamefont {I.}~\bibnamefont
  {{Gonz{\'{a}}lez-Adalid Pemart\'in}}}, \bibinfo {author} {\bibfnamefont
  {E.}~\bibnamefont {Momp{\'{o}}}}, \bibinfo {author} {\bibfnamefont
  {A.}~\bibnamefont {Lasanta}}, \bibinfo {author} {\bibfnamefont
  {V.}~\bibnamefont {Mart\'in-Mayor}},\ and\ \bibinfo {author} {\bibfnamefont
  {J.}~\bibnamefont {Salas}},\ }\bibfield  {title} {\bibinfo {title} {{Slow
  growth of magnetic domains helps fast evolution routes for out-of-equilibrium
  dynamics}},\ }\href@noop {} {\bibfield  {journal} {\bibinfo  {journal}
  {Physical Review E}\ }\textbf {\bibinfo {volume} {104}},\ \bibinfo {pages}
  {044114} (\bibinfo {year} {2021})}\BibitemShut {NoStop}%
\bibitem [{\citenamefont {Pemartin}\ \emph {et~al.}(2024)\citenamefont
  {Pemartin}, \citenamefont {Mompo}, \citenamefont {Lasanta}, \citenamefont
  {Martin-Mayor},\ and\ \citenamefont {Salas}}]{Pem24}%
  \BibitemOpen
  \bibfield  {author} {\bibinfo {author} {\bibfnamefont {I.~G.-A.}\
  \bibnamefont {Pemartin}}, \bibinfo {author} {\bibfnamefont {E.}~\bibnamefont
  {Mompo}}, \bibinfo {author} {\bibfnamefont {A.}~\bibnamefont {Lasanta}},
  \bibinfo {author} {\bibfnamefont {V.}~\bibnamefont {Martin-Mayor}},\ and\
  \bibinfo {author} {\bibfnamefont {J.}~\bibnamefont {Salas}},\ }\bibfield
  {title} {\bibinfo {title} {{Shortcuts of freely relaxing systems using
  equilibrium physical observables}},\ }\href@noop {} {\bibfield  {journal}
  {\bibinfo  {journal} {Physical Review Letters}\ }\textbf {\bibinfo {volume}
  {132}},\ \bibinfo {pages} {117102} (\bibinfo {year} {2024})}\BibitemShut
  {NoStop}%
\bibitem [{\citenamefont {Hwang}\ \emph {et~al.}(1993)\citenamefont {Hwang},
  \citenamefont {Hwang-Ma},\ and\ \citenamefont {Sheu}}]{Hwa93}%
  \BibitemOpen
  \bibfield  {author} {\bibinfo {author} {\bibfnamefont {C.-R.}\ \bibnamefont
  {Hwang}}, \bibinfo {author} {\bibfnamefont {S.-Y.}\ \bibnamefont
  {Hwang-Ma}},\ and\ \bibinfo {author} {\bibfnamefont {S.-J.}\ \bibnamefont
  {Sheu}},\ }\bibfield  {title} {\bibinfo {title} {{Accelerating gaussian
  diffusions}},\ }\href@noop {} {\bibfield  {journal} {\bibinfo  {journal} {The
  Annals of Applied Probability}\ }\textbf {\bibinfo {volume} {3}},\ \bibinfo
  {pages} {897} (\bibinfo {year} {1993})}\BibitemShut {NoStop}%
\bibitem [{\citenamefont {Suwa}\ and\ \citenamefont {Todo}(2010)}]{Suw10}%
  \BibitemOpen
  \bibfield  {author} {\bibinfo {author} {\bibfnamefont {H.}~\bibnamefont
  {Suwa}}\ and\ \bibinfo {author} {\bibfnamefont {S.}~\bibnamefont {Todo}},\
  }\bibfield  {title} {\bibinfo {title} {{Markov chain Monte Carlo method
  without detailed balance}},\ }\href@noop {} {\bibfield  {journal} {\bibinfo
  {journal} {Physical review letters}\ }\textbf {\bibinfo {volume} {105}},\
  \bibinfo {pages} {120603} (\bibinfo {year} {2010})}\BibitemShut {NoStop}%
\bibitem [{\citenamefont {Ichiki}\ and\ \citenamefont {Ohzeki}(2013)}]{Ich13}%
  \BibitemOpen
  \bibfield  {author} {\bibinfo {author} {\bibfnamefont {A.}~\bibnamefont
  {Ichiki}}\ and\ \bibinfo {author} {\bibfnamefont {M.}~\bibnamefont
  {Ohzeki}},\ }\bibfield  {title} {\bibinfo {title} {{Violation of detailed
  balance accelerates relaxation}},\ }\href@noop {} {\bibfield  {journal}
  {\bibinfo  {journal} {Physical Review E}\ }\textbf {\bibinfo {volume} {88}},\
  \bibinfo {pages} {20101} (\bibinfo {year} {2013})}\BibitemShut {NoStop}%
\bibitem [{\citenamefont {Coghi}\ \emph {et~al.}(2021)\citenamefont {Coghi},
  \citenamefont {Chetrite},\ and\ \citenamefont {Touchette}}]{Cog21}%
  \BibitemOpen
  \bibfield  {author} {\bibinfo {author} {\bibfnamefont {F.}~\bibnamefont
  {Coghi}}, \bibinfo {author} {\bibfnamefont {R.}~\bibnamefont {Chetrite}},\
  and\ \bibinfo {author} {\bibfnamefont {H.}~\bibnamefont {Touchette}},\
  }\bibfield  {title} {\bibinfo {title} {{Role of current fluctuations in
  nonreversible samplers}},\ }\href@noop {} {\bibfield  {journal} {\bibinfo
  {journal} {Physical Review E}\ }\textbf {\bibinfo {volume} {103}},\ \bibinfo
  {pages} {062142} (\bibinfo {year} {2021})}\BibitemShut {NoStop}%
\bibitem [{\citenamefont {Dieball}\ \emph {et~al.}(2023)\citenamefont
  {Dieball}, \citenamefont {Wellecke},\ and\ \citenamefont {Godec}}]{Die23}%
  \BibitemOpen
  \bibfield  {author} {\bibinfo {author} {\bibfnamefont {C.}~\bibnamefont
  {Dieball}}, \bibinfo {author} {\bibfnamefont {G.}~\bibnamefont {Wellecke}},\
  and\ \bibinfo {author} {\bibfnamefont {A.}~\bibnamefont {Godec}},\ }\bibfield
   {title} {\bibinfo {title} {{Asymmetric thermal relaxation in driven systems:
  Rotations go opposite ways}},\ }\href@noop {} {\bibfield  {journal} {\bibinfo
   {journal} {Physical Review Research}\ }\textbf {\bibinfo {volume} {5}},\
  \bibinfo {pages} {L042030} (\bibinfo {year} {2023})}\BibitemShut {NoStop}%
\bibitem [{\citenamefont {Martinez}\ \emph {et~al.}(2016)\citenamefont
  {Martinez}, \citenamefont {Petrosyan}, \citenamefont {Gu{\'{e}}ry-Odelin},
  \citenamefont {Trizac},\ and\ \citenamefont {Ciliberto}}]{Mar16}%
  \BibitemOpen
  \bibfield  {author} {\bibinfo {author} {\bibfnamefont {I.~A.}\ \bibnamefont
  {Martinez}}, \bibinfo {author} {\bibfnamefont {A.}~\bibnamefont {Petrosyan}},
  \bibinfo {author} {\bibfnamefont {D.}~\bibnamefont {Gu{\'{e}}ry-Odelin}},
  \bibinfo {author} {\bibfnamefont {E.}~\bibnamefont {Trizac}},\ and\ \bibinfo
  {author} {\bibfnamefont {S.}~\bibnamefont {Ciliberto}},\ }\bibfield  {title}
  {\bibinfo {title} {{Engineered swift equilibration of a Brownian particle}},\
  }\href@noop {} {\bibfield  {journal} {\bibinfo  {journal} {Nature Physics}\
  }\textbf {\bibinfo {volume} {12}},\ \bibinfo {pages} {843} (\bibinfo {year}
  {2016})}\BibitemShut {NoStop}%
\bibitem [{\citenamefont {Raynal}\ \emph {et~al.}(2023)\citenamefont {Raynal},
  \citenamefont {de~Guillebon}, \citenamefont {Gu{\'{e}}ry-Odelin},
  \citenamefont {Trizac}, \citenamefont {Lauret},\ and\ \citenamefont
  {Rondin}}]{Ray23}%
  \BibitemOpen
  \bibfield  {author} {\bibinfo {author} {\bibfnamefont {D.}~\bibnamefont
  {Raynal}}, \bibinfo {author} {\bibfnamefont {T.}~\bibnamefont
  {de~Guillebon}}, \bibinfo {author} {\bibfnamefont {D.}~\bibnamefont
  {Gu{\'{e}}ry-Odelin}}, \bibinfo {author} {\bibfnamefont {E.}~\bibnamefont
  {Trizac}}, \bibinfo {author} {\bibfnamefont {J.-S.}\ \bibnamefont {Lauret}},\
  and\ \bibinfo {author} {\bibfnamefont {L.}~\bibnamefont {Rondin}},\
  }\bibfield  {title} {\bibinfo {title} {{Shortcuts to Equilibrium with a
  Levitated Particle in the Underdamped Regime}},\ }\href
  {https://doi.org/10.1103/PhysRevLett.131.087101} {\bibfield  {journal}
  {\bibinfo  {journal} {Physical Review Letters}\ }\textbf {\bibinfo {volume}
  {131}},\ \bibinfo {pages} {087101} (\bibinfo {year} {2023})}\BibitemShut
  {NoStop}%
\bibitem [{\citenamefont {Nattermann}\ and\ \citenamefont
  {Vilfan}(1988)}]{Nat88}%
  \BibitemOpen
  \bibfield  {author} {\bibinfo {author} {\bibfnamefont {T.}~\bibnamefont
  {Nattermann}}\ and\ \bibinfo {author} {\bibfnamefont {I.}~\bibnamefont
  {Vilfan}},\ }\bibfield  {title} {\bibinfo {title} {{Anomalous Relaxation in
  the Random-Field Ising Model and Related Systems}},\ }\href
  {https://doi.org/10.1103/PhysRevLett.61.223} {\bibfield  {journal} {\bibinfo
  {journal} {Physical Review Letters}\ }\textbf {\bibinfo {volume} {61}},\
  \bibinfo {pages} {223} (\bibinfo {year} {1988})}\BibitemShut {NoStop}%
\bibitem [{\citenamefont {Cugliandolo}\ and\ \citenamefont
  {Kurchan}(1994)}]{Cug94}%
  \BibitemOpen
  \bibfield  {author} {\bibinfo {author} {\bibfnamefont {L.~F.}\ \bibnamefont
  {Cugliandolo}}\ and\ \bibinfo {author} {\bibfnamefont {J.}~\bibnamefont
  {Kurchan}},\ }\bibfield  {title} {\bibinfo {title} {{On the
  out-of-equilibrium relaxation of the Sherrington-Kirkpatrick model}},\
  }\href@noop {} {\bibfield  {journal} {\bibinfo  {journal} {Journal of Physics
  A: Mathematical and General}\ }\textbf {\bibinfo {volume} {27}},\ \bibinfo
  {pages} {5749} (\bibinfo {year} {1994})}\BibitemShut {NoStop}%
\bibitem [{\citenamefont {Klapp}\ and\ \citenamefont {Patey}(2001)}]{Kla01}%
  \BibitemOpen
  \bibfield  {author} {\bibinfo {author} {\bibfnamefont {S.~H.~L.}\
  \bibnamefont {Klapp}}\ and\ \bibinfo {author} {\bibfnamefont {G.~N.}\
  \bibnamefont {Patey}},\ }\bibfield  {title} {\bibinfo {title} {{Ferroelectric
  order in positionally frozen dipolar systems}},\ }\href@noop {} {\bibfield
  {journal} {\bibinfo  {journal} {The Journal of Chemical Physics}\ }\textbf
  {\bibinfo {volume} {115}},\ \bibinfo {pages} {4718} (\bibinfo {year}
  {2001})}\BibitemShut {NoStop}%
\bibitem [{\citenamefont {Woo}\ and\ \citenamefont {Monson}(2003)}]{Woo03}%
  \BibitemOpen
  \bibfield  {author} {\bibinfo {author} {\bibfnamefont {H.-J.}\ \bibnamefont
  {Woo}}\ and\ \bibinfo {author} {\bibfnamefont {P.~A.}\ \bibnamefont
  {Monson}},\ }\bibfield  {title} {\bibinfo {title} {{Phase behavior and
  dynamics of fluids in mesoporous glasses}},\ }\href@noop {} {\bibfield
  {journal} {\bibinfo  {journal} {Physical Review E}\ }\textbf {\bibinfo
  {volume} {67}},\ \bibinfo {pages} {041207} (\bibinfo {year}
  {2003})}\BibitemShut {NoStop}%
\bibitem [{\citenamefont {Detcheverry}\ \emph {et~al.}(2003)\citenamefont
  {Detcheverry}, \citenamefont {Kierlik}, \citenamefont {Rosinberg},\ and\
  \citenamefont {Tarjus}}]{Det03}%
  \BibitemOpen
  \bibfield  {author} {\bibinfo {author} {\bibfnamefont {F.}~\bibnamefont
  {Detcheverry}}, \bibinfo {author} {\bibfnamefont {E.}~\bibnamefont
  {Kierlik}}, \bibinfo {author} {\bibfnamefont {M.~L.}\ \bibnamefont
  {Rosinberg}},\ and\ \bibinfo {author} {\bibfnamefont {G.}~\bibnamefont
  {Tarjus}},\ }\bibfield  {title} {\bibinfo {title} {{Local mean-field study of
  capillary condensation in silica aerogels}},\ }\href@noop {} {\bibfield
  {journal} {\bibinfo  {journal} {Physical Review E}\ }\textbf {\bibinfo
  {volume} {68}},\ \bibinfo {pages} {061504} (\bibinfo {year}
  {2003})}\BibitemShut {NoStop}%
\bibitem [{\citenamefont {Biroli}(2015)}]{Bir15}%
  \BibitemOpen
  \bibfield  {author} {\bibinfo {author} {\bibfnamefont {G.}~\bibnamefont
  {Biroli}},\ }\bibfield  {title} {\bibinfo {title} {{Slow relaxations and
  non-equilibrium dynamics in classical and quantum systems}},\ }\href@noop {}
  {\bibfield  {journal} {\bibinfo  {journal} {arXiv preprint arXiv:1507.05858}\
  } (\bibinfo {year} {2015})}\BibitemShut {NoStop}%
\bibitem [{\citenamefont {Duan}\ \emph {et~al.}(2021)\citenamefont {Duan},
  \citenamefont {Mahault}, \citenamefont {Ma}, \citenamefont {Shi},\ and\
  \citenamefont {Chat{\'{e}}}}]{Dua21}%
  \BibitemOpen
  \bibfield  {author} {\bibinfo {author} {\bibfnamefont {Y.}~\bibnamefont
  {Duan}}, \bibinfo {author} {\bibfnamefont {B.}~\bibnamefont {Mahault}},
  \bibinfo {author} {\bibfnamefont {Y.-q.}\ \bibnamefont {Ma}}, \bibinfo
  {author} {\bibfnamefont {X.-q.}\ \bibnamefont {Shi}},\ and\ \bibinfo {author}
  {\bibfnamefont {H.}~\bibnamefont {Chat{\'{e}}}},\ }\bibfield  {title}
  {\bibinfo {title} {{Breakdown of ergodicity and self-averaging in polar
  flocks with quenched disorder}},\ }\href@noop {} {\bibfield  {journal}
  {\bibinfo  {journal} {Physical Review Letters}\ }\textbf {\bibinfo {volume}
  {126}},\ \bibinfo {pages} {178001} (\bibinfo {year} {2021})}\BibitemShut
  {NoStop}%
\bibitem [{\citenamefont {Dries}\ \emph {et~al.}(2010)\citenamefont {Dries},
  \citenamefont {Pollack}, \citenamefont {Hitchcock},\ and\ \citenamefont
  {Hulet}}]{Dri10}%
  \BibitemOpen
  \bibfield  {author} {\bibinfo {author} {\bibfnamefont {D.}~\bibnamefont
  {Dries}}, \bibinfo {author} {\bibfnamefont {S.~E.}\ \bibnamefont {Pollack}},
  \bibinfo {author} {\bibfnamefont {J.~M.}\ \bibnamefont {Hitchcock}},\ and\
  \bibinfo {author} {\bibfnamefont {R.~G.}\ \bibnamefont {Hulet}},\ }\bibfield
  {title} {\bibinfo {title} {{Dissipative transport of a Bose-Einstein
  condensate}},\ }\href {https://doi.org/10.1103/PhysRevA.82.033603} {\bibfield
   {journal} {\bibinfo  {journal} {Physical Review A}\ }\textbf {\bibinfo
  {volume} {82}},\ \bibinfo {pages} {033603} (\bibinfo {year}
  {2010})}\BibitemShut {NoStop}%
\bibitem [{\citenamefont {Volpe}\ \emph {et~al.}(2014)\citenamefont {Volpe},
  \citenamefont {Volpe},\ and\ \citenamefont {Gigan}}]{Vol14}%
  \BibitemOpen
  \bibfield  {author} {\bibinfo {author} {\bibfnamefont {G.}~\bibnamefont
  {Volpe}}, \bibinfo {author} {\bibfnamefont {G.}~\bibnamefont {Volpe}},\ and\
  \bibinfo {author} {\bibfnamefont {S.}~\bibnamefont {Gigan}},\ }\bibfield
  {title} {\bibinfo {title} {{Brownian motion in a speckle light field: tunable
  anomalous diffusion and selective optical manipulation}},\ }\href@noop {}
  {\bibfield  {journal} {\bibinfo  {journal} {Scientific Reports}\ }\textbf
  {\bibinfo {volume} {4}},\ \bibinfo {pages} {3936} (\bibinfo {year}
  {2014})}\BibitemShut {NoStop}%
\bibitem [{\citenamefont {Hanes}\ \emph {et~al.}(2012)\citenamefont {Hanes},
  \citenamefont {Dalle-Ferrier}, \citenamefont {Schmiedeberg}, \citenamefont
  {Jenkins},\ and\ \citenamefont {Egelhaaf}}]{Han12}%
  \BibitemOpen
  \bibfield  {author} {\bibinfo {author} {\bibfnamefont {R.~D.~L.}\
  \bibnamefont {Hanes}}, \bibinfo {author} {\bibfnamefont {C.}~\bibnamefont
  {Dalle-Ferrier}}, \bibinfo {author} {\bibfnamefont {M.}~\bibnamefont
  {Schmiedeberg}}, \bibinfo {author} {\bibfnamefont {M.~C.}\ \bibnamefont
  {Jenkins}},\ and\ \bibinfo {author} {\bibfnamefont {S.~U.}\ \bibnamefont
  {Egelhaaf}},\ }\bibfield  {title} {\bibinfo {title} {{Colloids in one
  dimensional random energy landscapes}},\ }\href@noop {} {\bibfield  {journal}
  {\bibinfo  {journal} {Soft Matter}\ }\textbf {\bibinfo {volume} {8}},\
  \bibinfo {pages} {2714} (\bibinfo {year} {2012})}\BibitemShut {NoStop}%
\bibitem [{\citenamefont {Bewerunge}\ and\ \citenamefont
  {Egelhaaf}(2016)}]{Bew16a}%
  \BibitemOpen
  \bibfield  {author} {\bibinfo {author} {\bibfnamefont {J.}~\bibnamefont
  {Bewerunge}}\ and\ \bibinfo {author} {\bibfnamefont {S.~U.}\ \bibnamefont
  {Egelhaaf}},\ }\bibfield  {title} {\bibinfo {title} {{Experimental creation
  and characterization of random potential-energy landscapes exploiting speckle
  patterns}},\ }\href@noop {} {\bibfield  {journal} {\bibinfo  {journal}
  {Physical Review A}\ }\textbf {\bibinfo {volume} {93}},\ \bibinfo {pages}
  {013806} (\bibinfo {year} {2016})}\BibitemShut {NoStop}%
\bibitem [{\citenamefont {Zunke}\ \emph {et~al.}(2022)\citenamefont {Zunke},
  \citenamefont {Bewerunge}, \citenamefont {Platten}, \citenamefont
  {Egelhaaf},\ and\ \citenamefont {Godec}}]{Zun22}%
  \BibitemOpen
  \bibfield  {author} {\bibinfo {author} {\bibfnamefont {C.}~\bibnamefont
  {Zunke}}, \bibinfo {author} {\bibfnamefont {J.}~\bibnamefont {Bewerunge}},
  \bibinfo {author} {\bibfnamefont {F.}~\bibnamefont {Platten}}, \bibinfo
  {author} {\bibfnamefont {S.~U.}\ \bibnamefont {Egelhaaf}},\ and\ \bibinfo
  {author} {\bibfnamefont {A.}~\bibnamefont {Godec}},\ }\bibfield  {title}
  {\bibinfo {title} {{First-passage statistics of colloids on fractals: Theory
  and experimental realization}},\ }\href@noop {} {\bibfield  {journal}
  {\bibinfo  {journal} {Science advances}\ }\textbf {\bibinfo {volume} {8}},\
  \bibinfo {pages} {eabk0627} (\bibinfo {year} {2022})}\BibitemShut {NoStop}%
\bibitem [{\citenamefont {Franz}\ and\ \citenamefont
  {M{\'{e}}zard}(1994)}]{Fra94}%
  \BibitemOpen
  \bibfield  {author} {\bibinfo {author} {\bibfnamefont {S.}~\bibnamefont
  {Franz}}\ and\ \bibinfo {author} {\bibfnamefont {M.}~\bibnamefont
  {M{\'{e}}zard}},\ }\bibfield  {title} {\bibinfo {title} {{Off-equilibrium
  glassy dynamics: a simple case}},\ }\href@noop {} {\bibfield  {journal}
  {\bibinfo  {journal} {Europhysics Letters}\ }\textbf {\bibinfo {volume}
  {26}},\ \bibinfo {pages} {209} (\bibinfo {year} {1994})}\BibitemShut
  {NoStop}%
\bibitem [{\citenamefont {Bhongale}\ \emph {et~al.}(2010)\citenamefont
  {Bhongale}, \citenamefont {Kakashvili}, \citenamefont {Bolech},\ and\
  \citenamefont {Pu}}]{Bho10}%
  \BibitemOpen
  \bibfield  {author} {\bibinfo {author} {\bibfnamefont {S.~G.}\ \bibnamefont
  {Bhongale}}, \bibinfo {author} {\bibfnamefont {P.}~\bibnamefont
  {Kakashvili}}, \bibinfo {author} {\bibfnamefont {C.~J.}\ \bibnamefont
  {Bolech}},\ and\ \bibinfo {author} {\bibfnamefont {H.}~\bibnamefont {Pu}},\
  }\bibfield  {title} {\bibinfo {title} {{Dissipative transport of trapped
  Bose-Einstein condensates through disorder}},\ }\href@noop {} {\bibfield
  {journal} {\bibinfo  {journal} {Physical Review A}\ }\textbf {\bibinfo
  {volume} {82}},\ \bibinfo {pages} {053632} (\bibinfo {year}
  {2010})}\BibitemShut {NoStop}%
\bibitem [{\citenamefont {Hsueh}\ \emph {et~al.}(2018)\citenamefont {Hsueh},
  \citenamefont {Ong}, \citenamefont {Tseng}, \citenamefont {Tsubota},\ and\
  \citenamefont {Wu}}]{Hsu18}%
  \BibitemOpen
  \bibfield  {author} {\bibinfo {author} {\bibfnamefont {C.-H.}\ \bibnamefont
  {Hsueh}}, \bibinfo {author} {\bibfnamefont {R.}~\bibnamefont {Ong}}, \bibinfo
  {author} {\bibfnamefont {J.-F.}\ \bibnamefont {Tseng}}, \bibinfo {author}
  {\bibfnamefont {M.}~\bibnamefont {Tsubota}},\ and\ \bibinfo {author}
  {\bibfnamefont {W.-C.}\ \bibnamefont {Wu}},\ }\bibfield  {title} {\bibinfo
  {title} {{Thermalization and localization of an oscillating Bose-Einstein
  condensate in a disordered trap}},\ }\href
  {https://doi.org/10.1103/PhysRevA.98.063613} {\bibfield  {journal} {\bibinfo
  {journal} {Physical Review A}\ }\textbf {\bibinfo {volume} {98}},\ \bibinfo
  {pages} {063613} (\bibinfo {year} {2018})}\BibitemShut {NoStop}%
\bibitem [{\citenamefont {Hsueh}\ \emph {et~al.}(2020)\citenamefont {Hsueh},
  \citenamefont {Hsueh},\ and\ \citenamefont {Wu}}]{Hsu20}%
  \BibitemOpen
  \bibfield  {author} {\bibinfo {author} {\bibfnamefont {Y.-W.}\ \bibnamefont
  {Hsueh}}, \bibinfo {author} {\bibfnamefont {C.-H.}\ \bibnamefont {Hsueh}},\
  and\ \bibinfo {author} {\bibfnamefont {W.-C.}\ \bibnamefont {Wu}},\
  }\bibfield  {title} {\bibinfo {title} {{Thermalization in a quantum harmonic
  oscillator with random disorder}},\ }\href@noop {} {\bibfield  {journal}
  {\bibinfo  {journal} {Entropy}\ }\textbf {\bibinfo {volume} {22}},\ \bibinfo
  {pages} {855} (\bibinfo {year} {2020})}\BibitemShut {NoStop}%
\bibitem [{\citenamefont {Sch{\"{u}}rger}\ \emph {et~al.}(2022)\citenamefont
  {Sch{\"{u}}rger}, \citenamefont {Schaupp}, \citenamefont {Kaiser},
  \citenamefont {Engels},\ and\ \citenamefont {Engel}}]{Schu22}%
  \BibitemOpen
  \bibfield  {author} {\bibinfo {author} {\bibfnamefont {P.}~\bibnamefont
  {Sch{\"{u}}rger}}, \bibinfo {author} {\bibfnamefont {T.}~\bibnamefont
  {Schaupp}}, \bibinfo {author} {\bibfnamefont {D.}~\bibnamefont {Kaiser}},
  \bibinfo {author} {\bibfnamefont {B.}~\bibnamefont {Engels}},\ and\ \bibinfo
  {author} {\bibfnamefont {V.}~\bibnamefont {Engel}},\ }\bibfield  {title}
  {\bibinfo {title} {{Wave packet dynamics in an harmonic potential disturbed
  by disorder: Entropy, uncertainty, and vibrational revivals}},\ }\href@noop
  {} {\bibfield  {journal} {\bibinfo  {journal} {The Journal of Chemical
  Physics}\ }\textbf {\bibinfo {volume} {156}} (\bibinfo {year}
  {2022})}\BibitemShut {NoStop}%
\bibitem [{\citenamefont {Risken}(1989)}]{Ris89}%
  \BibitemOpen
  \bibfield  {author} {\bibinfo {author} {\bibfnamefont {H.}~\bibnamefont
  {Risken}},\ }\href@noop {} {\emph {\bibinfo {title} {{The Fokker-Planck
  Equation: Methods of Solution and Applications, 2nd edition}}}}\ (\bibinfo
  {publisher} {Springer},\ \bibinfo {address} {Berlin, Germany},\ \bibinfo
  {year} {1989})\BibitemShut {NoStop}%
\bibitem [{Note1()}]{Note1}%
  \BibitemOpen
  \bibinfo {note} {The intensity of laser speckles is typically Gaussian
  correlated~\cite {Goo07}. Therefore, Gaussian correlation functions are
  relevant for speckle-perturbed laser traps.}\BibitemShut {Stop}%
\bibitem [{Note2()}]{Note2}%
  \BibitemOpen
  \bibinfo {note} {As we explain later, $\langle \Delta \lambda _1\rangle _V$
  behaves as $\langle \Delta \lambda _1\rangle _V\propto \zeta _T^2$ for $\zeta
  _T\ll 1$. Therefore, the change of sign of $\langle \Delta \lambda _1\rangle
  _V$ is only marginally reflected by the colour coding in Fig.~\ref
  {fig:phasediag}(a) for small $\zeta _T$.}\BibitemShut {Stop}%
\bibitem [{sup()}]{supp}%
  \BibitemOpen
  \href@noop {} {\bibinfo {title} {{See Supplemental Material for the mean and
  variance of $\Delta\lambda_1$ associated with the sinc and exponential
  correlation functions.}}}\BibitemShut {Stop}%
\bibitem [{\citenamefont {Davydov}(1976)}]{Dav76}%
  \BibitemOpen
  \bibfield  {author} {\bibinfo {author} {\bibfnamefont {A.~S.}\ \bibnamefont
  {Davydov}},\ }\href@noop {} {\emph {\bibinfo {title} {{Quantum
  Mechanics}}}},\ \bibinfo {edition} {ii}\ ed.\ (\bibinfo  {publisher}
  {Pergamon Press Ltd.},\ \bibinfo {address} {Oxford, UK},\ \bibinfo {year}
  {1976})\BibitemShut {NoStop}%
\bibitem [{Note3()}]{Note3}%
  \BibitemOpen
  \bibinfo {note} {We use the convention $\protect \hat f(k)=\DOTSI \intop
  \ilimits@ _{-\infty }^{\infty }\protect \!\protect \!\protect \ensuremath
  {\protect \text {d}}{x}\protect \ensuremath {\protect \text
  {e}}^{-ikx}f(x)/(2\pi )$ for the Fourier transform of $f$.}\BibitemShut
  {Stop}%
\bibitem [{dlm()}]{dlmf}%
  \BibitemOpen
  \href {http://dlmf.nist.gov/} {\bibinfo {title} {{\it NIST Digital Library of
  Mathematical Functions}}},\ \bibinfo {howpublished}
  {{\url{http://dlmf.nist.gov/}}, Release 1.1.2 of 2021-06-15}\BibitemShut
  {NoStop}%
\bibitem [{\citenamefont {M{\'{e}}zard}\ \emph {et~al.}(1987)\citenamefont
  {M{\'{e}}zard}, \citenamefont {Parisi},\ and\ \citenamefont
  {Virasoro}}]{Mez87}%
  \BibitemOpen
  \bibfield  {author} {\bibinfo {author} {\bibfnamefont {M.}~\bibnamefont
  {M{\'{e}}zard}}, \bibinfo {author} {\bibfnamefont {G.}~\bibnamefont
  {Parisi}},\ and\ \bibinfo {author} {\bibfnamefont {M.~A.}\ \bibnamefont
  {Virasoro}},\ }\href@noop {} {\emph {\bibinfo {title} {{Spin Glass Theory and
  Beyond: An Introduction to the Replica Method and Its Applications}}}},\
  Vol.~\bibinfo {volume} {9}\ (\bibinfo  {publisher} {World Scientific
  Publishing Company},\ \bibinfo {address} {Singapore},\ \bibinfo {year}
  {1987})\BibitemShut {NoStop}%
\bibitem [{\citenamefont {Bender}\ and\ \citenamefont {Orszag}(1978)}]{Ben78}%
  \BibitemOpen
  \bibfield  {author} {\bibinfo {author} {\bibfnamefont {C.~M.}\ \bibnamefont
  {Bender}}\ and\ \bibinfo {author} {\bibfnamefont {S.~A.}\ \bibnamefont
  {Orszag}},\ }\href@noop {} {\emph {\bibinfo {title} {{Advanced Mathematical
  Methods for Scientists and Engineers}}}}\ (\bibinfo  {publisher}
  {McGraw-Hill},\ \bibinfo {address} {New York, USA},\ \bibinfo {year}
  {1978})\BibitemShut {NoStop}%
\bibitem [{\citenamefont {Trefethen}\ and\ \citenamefont
  {Embree}(2005)}]{Tre05}%
  \BibitemOpen
  \bibfield  {author} {\bibinfo {author} {\bibfnamefont {L.~N.}\ \bibnamefont
  {Trefethen}}\ and\ \bibinfo {author} {\bibfnamefont {M.}~\bibnamefont
  {Embree}},\ }\href@noop {} {\emph {\bibinfo {title} {{Spectra and
  Pseudospectra: The Behavior of Nonnormal Matrices and Operators}}}}\
  (\bibinfo  {publisher} {Princeton University Press},\ \bibinfo {address}
  {Princeton, NJ},\ \bibinfo {year} {2005})\BibitemShut {NoStop}%
\bibitem [{\citenamefont {Goodman}(2007)}]{Goo07}%
  \BibitemOpen
  \bibfield  {author} {\bibinfo {author} {\bibfnamefont {J.~W.}\ \bibnamefont
  {Goodman}},\ }\href@noop {} {\emph {\bibinfo {title} {{Speckle Phenomena in
  Optics: Theory and Applications}}}}\ (\bibinfo  {publisher} {Roberts and
  Company Publishers},\ \bibinfo {address} {Greenwood Village, CO},\ \bibinfo
  {year} {2007})\BibitemShut {NoStop}%
\end{thebibliography}
\end{document}